\documentclass[a4paper,11pt]{article}
\pdfoutput=1 

\usepackage{jheppub} 

\usepackage[T1]{fontenc} 
\usepackage{amssymb,epsf}
\usepackage{amsmath,amssymb}
\usepackage{graphicx}
\usepackage{setspace}
\usepackage{fancyhdr}
\usepackage{ifpdf}
\usepackage{graphicx}
\usepackage{color}
\usepackage{comment}
\usepackage{wrapfig}

\usepackage{xcolor}

\newcommand{\Slash}[1]{{\ooalign{\hfil$#1$\hfil\crcr\raise.167ex\hbox{/}}}}
\newcommand{\be}{\begin{equation}}
\newcommand{\ba}{\begin{eqnarray}}
\newcommand{\ea}{\end{eqnarray}}
\newcommand{\ee}{\end{equation}}

\def\abs#1{\mid \! #1 \! \mid}

\def\see{S^{EE}_A}

\numberwithin{equation}{section}
\allowdisplaybreaks  

\title{\boldmath Holographic entanglement entropy of a $1+1$ dimensional $p$-wave superconductor}

\author[a]{Sumit R. Das,}
\author[a]{Mitsutoshi Fujita} 
\author[a]{and Bom Soo Kim}

\affiliation[a]{Department of Physics and Astronomy, University of Kentucky,
Lexington, KY 40506, USA}

\emailAdd{das@pa.uky.edu}
\emailAdd{mitsutoshi.fujita@uky.edu}
\emailAdd{bom.soo.kim@uky.edu}

\abstract{We examine the behavior of entanglement entropy $\see$ of a subsystem $A$ in a fully backreacted holographic model of a $1+1$ dimensional $p$ wave superconductor across the phase transition. For a given temperature, the system goes to a superconducting phase beyond a critical value of the charge density. The entanglement entropy, considered as a function of the charge density at a given temperature, has a cusp at the critical point. In addition, we find that there are three different behaviors in the condensed phase, depending on the subsystem size. 
For a subsystem size $l$ smaller than a critical size $l_{c1}$,  $\see$ continues to increase as a function of the charge density as we cross the phase transition. When $l$ lies between $l_{c1}$ and another critical size $l_{c2}$ the entanglement entropy displays a non-monotonic behavior, while for $l > l_{c2}$ it decreases monotonically. At large charge densities $\see$ appears to saturate. The non-monotonic behavior leads to a novel phase diagram for this system.}

\begin{document} 
\maketitle
\flushbottom

\section{Introduction}

The entanglement entropy $\see$ of a subsystem $A$ of a system is a useful non-local quantity in quantum field theories ~\cite{EE,Cardy,Review}. This quantity often provides interesting probes of the physics of phase transitions, most notably for quantum critical transitions where it typically diverges \cite{Vidal:2002rm}. Even for thermal transitions, the entanglement entropy provides additional diagnostics of the nature of the phases, since it is sensitive to the number of available degrees of freedom. In recent years, the Ryu-Takayanagi prescription has become a key tool to compute entanglement entropy of strongly coupled systems which have gravity duals ~\cite{Ryu:2006bv,Ryu:2006ef,Nishioka:2009un}. Indeed, starting with \cite{Albash:2012pd}, there have been several studies of the change of behavior of the entanglement entropy in holographic models of critical phase transitions. \cite{Albash:2012pd} examined a model of s-wave holographic superconducting transition,~\cite{Cai:2012sk} considered a model of an insulator-superfluid transition while \cite{Cai:2012nm} studies a model of holographic $p$-wave superconductor phase transition based on \cite{Gubser:2008wv,Ammon:2009xh}. Recently, \cite{Kuang:2014kha,Zangeneh:2017tub} extended the study of the holographic entanglement entropy near critical phase transitions. In all these models a suitable order parameter condenses in the superfluid phase, reducing the number of degrees of freedom.  Indeed these studies found that the magnitude of the entropy in the condensed phase is always lower than what it would have been in the absence of condensation.
In all these studies, the entanglement entropy as a function of the temperature or the chemical potential displayed a cusp at the location of the transition where the relevant derivative becomes discontinuous. As one goes into the condensed phase the entanglement entropy sometimes shows a non-monotonic behavior close to the transition, while far from the transition it keeps decreasing.

In this paper, we perform a calculation of the holographic entanglement entropy (HEE) in a toy model of a $1+1$ dimensional $p$-wave superconductor at non-zero charge density and temperature. The dual theory is a $2+1$ theory of gravity and $SU(2)$ Yang-Mills field with a negative cosmological constant. The action is given by
\ba\label{GRE1}
I_{G}=\dfrac{1}{2\kappa^2}\int d^3x\sqrt{-g}\Big(R+\dfrac{2}{L^2}-\mbox{Tr}(\tilde{F}_{\mu\nu}\tilde{F}^{\mu\nu})\Big),
\ea
where $A_\mu$ is a $SU(2)$ gauge field with the field strength $\tilde{F}_{\mu\nu}=\partial_{\mu}\tilde{A}_{\nu}-\partial_{\nu}\tilde{A}_{\mu}-ig_{YM}[\tilde{A}_{\mu},\tilde{A}_{\nu}]$. In this convention, $\tilde{A}_{\mu}$ is dimensionless, while $g_{YM}$ and $1/\kappa^2$ have the dimension 1. This model has been introduced in \cite{Gao:2012yw} following earlier work in higher dimensions \cite{Gubser:2008wv}. An important well known aspect of this $1+1$ dimensional model is that the gauge field  has to be treated in alternative quantization \cite{Jensen:2010em} : the non-normalizable part of the field can be interpreted as the expectation value of a {\em dynamical gauge field} living on the boundary, which makes it morally closer to true superconductors, as opposed to superfluids.

In the normal phase the dual geometry is a $AdS_3$ charged BTZ black hole, with the role of the usual Maxwell field played by the time component of the diagonal gauge field $A_t^{(3)}$. The boundary theory then has a nonzero temperature and charge density. As we increase the charge density $\rho$ for a given temperature, there is a critical value $\rho_c$ beyond which the black hole acquires a vector hair, e.g. a spatial component $A_x^{(1)}$ becomes nonzero \footnote{This is different from vector hair in addition to a scalar condensate which represents a current \cite{basu}.} : its boundary value then becomes the expectation value of a vector order parameter in the dual field theory. In the probe approximation the corresponding bulk solutions have been obtained in \cite{Gao:2012yw}. 

The dual field theory lives on a circle with circumference $2\pi L$ (where $L$ is the AdS scale as in equation (\ref{GRE1}), and we calculate the holographic entanglement entropy of an interval of size $l$ in this circle. In the normal phase the exact gravity solution is of course known. In the condensed phase the solution is known only in the probe approximation. To obtain the holographic entanglement entropy we first obtain the fully backreacted solution for the hairy charged BTZ black hole by numerically integrating the bulk equations of motion. Using this backreacted solution we compute the Bekenstein-Hawking entropy $S_{BH} = \frac{A}{2\kappa^2}$ (where $A$ is the horizon area and $\kappa$ is the gravitational constant) and the entanglement entropy $S_{EE}$ using the Ryu-Takayanagi prescription
\ba 
S_{EE}=\dfrac{2\pi}{\kappa^2}(\gamma_A),
\ea
where $\gamma_A$ is the length of the one-dimensional bulk geodesic whose end points coincide with the two end points of the interval on the boundary.  We focus on the behavior of these quantities at some fixed temperature as a function of the charge density for various values of $l$. 

We find that the Bekenstein-hawking entropy $S_{BH}$ has a cusp at the critical charge density $q=q_c$ where $\frac{\partial S_{BH}}{\partial q}$ is discontinuous. For $q > q_c$ this monotonically decreases, approaching a constant value. This decrease is essentially due to the reduction of degrees of freedom due to condensation.

For a given temperature and $q < q_c$ the entanglement entropy monotonically increases as a function of $q$ and its derivative has a discontinuity at $q = q_c$.  This cusp-like behavior has been observed earlier in higher dimensional models. However the behavior for $q > q_c$ appears quite different from that in the higher dimensional examples. This depends on the subsystem size $l$ and there are two critical sizes $l_{c1},l_{c2}$ where the behavior changes. For small enough size, $l < l_{c1}$ the entanglement entropy keeps increasing monotonically. When $l_{c2} < l < l_{c1}$ this quantity displays a non-monotonic behavior, decreasing at first to a local minimum, and then increasing to a maximum value and finally decreasing again. The local minimum moves closer to $\rho = \rho_c$ as $l$ approaches $l_{c1}$. At $l=l_{c2}$ the local minima and maxima merge to a point of inflection, so that for $l > l_{c2}$ the entanglement entropy decreases monotonically with increasing charge density.

While we do not have a complete understanding, it appears that this complex behavior arises from a competition between charge density and condensation. If there was no condensation, the entanglement entropy would have kept increasing with charge density for any subsystem size. Condensation reduces the number of effective degrees of freedom. When the subsystem size is small enough, the effects of condensation are small, and the tendency of the entanglement entropy to increase with charge density wins over. When the subsystem size is large enough, the effect of condensation wins over. The non-monotonic behavior seen for intermediate sizes reflects a competition between these two effects. This results in an interesting "phase diagram" where the sign of $(\partial S_{EE} / \partial q)_l$ distinguishes different "phases".

In Section (\ref{sec:one}) we describe the system and the associated critical transition, and calculate the backreaction by numerically solving the equations of motion. The backreacted solution is then used to calculate the Bekenstein-Hawking entropy. In Section (\ref{sec:two}) we use the solution above to calculate the minimal geodesics and hence the holographic entanglement entropy. (\ref{sec:discussion}) contains a brief discussion.

\section{The backreacted solution} \label{sec:one}
  
The action of Einstein-Yang Mills system we consider in this paper is
\ba\label{GRE1}
I_{G}=\dfrac{1}{2\kappa^2}\int d^3x\sqrt{-g}\Big(R+\dfrac{2}{\tilde L^2}-\mbox{Tr}(\tilde{F}_{\mu\nu}\tilde{F}^{\mu\nu})\Big),
\ea
where the field strength of the $SU(2)$ Yang-Mills term is given by $\tilde{F}_{\mu\nu}=\partial_{\mu}\tilde{A}_{\nu}-\partial_{\nu}\tilde{A}_{\mu}-ig_{YM}[\tilde{A}_{\mu},\tilde{A}_{\nu}]$. One of the spatial directions (called $y$ below) is compact, $y\sim y+2\pi L$. In (\ref{GRE1}), $\tilde{A}_{\mu}$ is dimensionless, while $g_{YM}$ and $1/\kappa^2$ have the mass dimension 1. A standard field redefinition $\tilde{A}_{\mu}\to A_{\mu}/g_{YM}$ the field strength becomes ${F}_{\mu\nu}=\partial_{\mu}{A}_{\nu}-\partial_{\nu}{A}_{\mu}-i[{A}_{\mu},{A}_{\nu}]$ while the Yang-Mills kinetic term becomes proportional to $1/(\kappa^2g_{YM}^2)$. In this section, we analyze the backreaction (whose strength is controlled by $1/g_{YM}^2$) of the $SU(2)$ Yang-Mills term into the metric of the 2+1 dimensional gravity to compute the holographic entanglement entropy.  

The equations of motion of the gauge field and the metric derived from the above action become  
\ba\label{EOMG}
&D_{\mu}(\sqrt{-g}F^{\mu\nu})= \partial_{\mu}(\sqrt{-g}F^{\mu\nu})-i\sqrt{-g}[A_{\mu},F^{\mu\nu}]=0,  \\
&R_{\mu\nu}-\dfrac{1}{2}g_{\mu\nu}\Big(R+\dfrac{2}{\tilde L^2}\Big)=\kappa^2 T_{\mu\nu}, \label{EIN212}
\ea
where the energy momentum tensor is 
\ba\label{ENE213}
& T_{\mu\nu}=\dfrac{2}{\kappa^2}\mbox{tr}\Big(\tilde{F}_{\mu\alpha}\tilde{F}_{\nu}{}^{\alpha}-\dfrac{1}{4}g_{\mu\nu} \tilde{F}_{\alpha\beta}\tilde{F}^{\alpha\beta}\Big) 
=\dfrac{2}{\kappa^2 g_{YM}^2}\mbox{tr}\Big(F_{\mu\alpha}F_{\nu}{}^{\alpha}-\dfrac{1}{4}g_{\mu\nu} F_{\alpha\beta}F^{\alpha\beta}\Big). 
\ea 
If we set the Yang-Mills field to zero, the solutions to Einstein equations are either pure $AdS_3$ or an uncharged BTZ black hole.

In the presence of a nontrivial gauge field which is assumed to depend only on the radial $AdS$ direction, the Einstein equations \eqref{EOMG} and \eqref{EIN212} can be solved with an ansatz :
\ba\label{ANS13}
&ds^2=\dfrac{\tilde L^2}{z^2}\Big(-f_2(z)dt^2+dy^2+\dfrac{dz^2}{h_2(z)f_2(z)}\Big),  \\
&A=\dfrac{1}{2}\Big(\phi (z)  \sigma^3 dt+w(z) \cdot  \sigma^1 dy\Big),\quad A_z^b=0, \nonumber
\ea
where $\sigma^a\ (a=1,2,3)$ are the Pauli matrices.
The function $f_2(z)$ vanishes at $z=z_h$, the location of the black hole horizon.

The Einstein equations can be re-written as the following three equations:
\ba\label{EIN245}
&&\dfrac{f_2 \left(z h_2 f_2'+z f_2 h_2'-2 f_2 h_2+2\right)}{ z^2}-\dfrac{z^2 (\phi^2 w^2 + f_2 h_2 (\phi^{\prime 2} + f_2 w^{\prime 2}))}{g_{YM}^2 \tilde L^2}=0,  \\
&&\dfrac{2 z^2 h_2 f_2''+z^2 f_2' h_2'-4 z h_2 f_2'-2 z f_2 h_2'+4 f_2 h_2-4}{2 z^2 } 
+ \dfrac{z^2 \left(\phi^2 w^2-f_2 h_2 \left(f_2 w^{\prime 2}+\phi^{\prime 2}\right)\right)}{g_{YM}^2 \tilde L^2 f_2}=0, \nonumber \\
&&-\dfrac{z h_2 f_2'-2 f_2h_2+2}{ z^2 f_2 h_2}- \dfrac{ z^2 \left(f_2 h_2 \left(f_2 w^{\prime 2}-\phi^{\prime 2}\right)+\phi^2 w^2\right)}{g_{YM}^2 \tilde L^2 f_2^2 h_2}=0, \nonumber
\ea
The last equation is a constraint.
Using the ansatz the equations of motion for the Yang-Mills fields become
\ba\label{EOM37}
&-\sqrt{h_2}f_2(z\sqrt{h_2} \phi')'+zw^2 \phi =0, \\
&\sqrt{h_2}f_2(z \sqrt{h_2}f_2w')'+z \phi^2 w=0.  \nonumber 
\ea
Note that the equations of motion depend only on the product of $AdS$ radius $\tilde L$ and $g_{YM}$. This can be used to set the scale $\tilde L=1$. 

In the uncondensed phase the $y$ component of the gauge field $w(z)$ vanishes. In this case, nonlinear terms are absent in the field strength and the system reduces to an Einstein-Maxwell system. The solution is then well known : it is a charged $AdS_3$ black hole considered in \cite{Banados:1992wn,Jensen:2010em} and described in \eqref{EIN245} with 
\ba\label{btz230}
f_2(z)=1-\Big(\dfrac{z}{z_0}\Big)^2+\dfrac{q^2}{g_{YM}^2} z^2\log \Big(\dfrac{z}{z_0}\Big),\quad  h_2(z)=1, \quad \phi(z) = q\log \Big(\dfrac{z}{z_0}\Big).
\ea
The black hole horizon is located at $z=z_0$. The Hawking temperature is given by 
\ba
T_H=\dfrac{1}{4\pi}\Big(\dfrac{2}{z_0}-\dfrac{q^2z_0}{g_{YM}^2}\Big).
\ea
Furthermore, we have chosen $\phi(z)$ to vanish at the horizon, as required by regularity of the solution in the euclidean domain.
The regularized mass of the black hole, $M_0=(L/z_0)^2$, is inversely proportional to the squared horizon position and satisfies the BPS-like bound $M_0\ge (q/\sqrt{2}g_{YM})^2$~\cite{Cadoni:2009bn} which is saturated at $T_H=0$. At the equality, the horizon position $z_h$ is equal to the value of $z$ where $f(z)$ has its extremal value.

As explained in \cite{Jensen:2010em} the presence of a logarithmic term implies that the gauge field has to be treated in alternative quantization, which requires us to fix boundary conditions which specify the charge density $q$. The chemical potential is then given by the constant part $\mu = -q \log z_0$. The mass dimension of $q$ and $\mu$ are both $1$.

When the charge density is large enough there is another solution which has $w(z)\neq 0$, with a free energy lower than a charged $AdS_3$ black hole. 

The behavior of this solution near the $AdS$ boundary $z=0$ is
\ba\label{BOU347}
&&\phi(z)\sim q \log \Big(\dfrac{z}{z_p}\Big), \\
&&w(z)\sim w_c+J_w\log (z), \nonumber \\
&&f_2(z)\sim \dfrac{1-r_0^2 z^2}{h_0}+\dfrac{ q^2}{g_{YM}^2} z^2 \log (z), \nonumber\\
&&h_2(z)\sim h_0, \nonumber
\ea
where once again $q$ is the charge density of the boundary field theory while the chemical potential is given by $\mu\equiv -q\log (z_p)$
\footnote{$z_p$ inside log is normalized by $L$ to be dimensionless~\cite{Ejaz:2013fla}.}. 
The parameter $w_c$ is now interpreted as the expectation value of a vector order parameter in the boundary theory while $J_w$ is the source for this order parameter. Both these parameters have mass dimension 1. $h_0$ and $r_0$ are constants.

The black hole horizon is now located at $z=z_h$ as specified by the condition $f_2(z_h)=0$. Regularity at the horizon requires as usual $\phi (z_h)=0$. The fields can be now expanded near the horizon as follows: 
\ba\label{EXP210}
&&\phi(z)=a_1(z_h-z)+\dots, \\
&&w(z)={b_1}+b_2(z_h-z)+\dots, \nonumber \\
&&f_2(z)=d_2 (z_h-z)+\dots, \nonumber \\
&&h_2(z)={c_1}+c_2(z_h-z)+\dots . \nonumber   
\ea
The coefficients $(a_1, b_1, b_2, c_1, c_2, d_2)$ are constants.
In terms of these parameters the Hawking temperature is
\ba
T_{H}=\dfrac{1}{4\pi}\abs{f_2'(z_h)}\sqrt{h_2(z_h)}=\dfrac{\abs{d_2}\sqrt{c_1}}{4\pi}.
\ea

Substitution of \eqref{EXP210} into equations of motion \eqref{EIN245} and \eqref{EOM37} yields four independent parameters $(a_1,b_1,c_1,z_h)$. Other parameters like $b_2, c_2$ and $d_2$ are then represented by these 4 parameters and $g_{YM}$: $b_2=0$, $d_2={2}/{(c_1z_h)}-{a_1^2 z_h^3}/{g_{YM}^2}$, and $c_2=-2(a_1b_1c_1)^2z_h^5g_{YM}^2/(a_1^2 c_1z_h^4-2g_{YM}^2)^2$.

We now obtain a backreacted solution by numerically solving the equations of motion. Recall that the last equation of \eqref{EIN245} is a constraint equation. This can be solved by choosing regularity conditions at the horizon. We then solve the first two Einstein equations of \eqref{EIN245} and two equations \eqref{EOM37} numerically, starting from the horizon and proceeding to the $AdS$ boundary. 

To look for a superconducting phase we need to find solutions with a vanishing source in the boundary theory. This implies that at the $AdS$ boundary, $J_w$ has to vanish. A nontrivial solution for $w(z)$ then signifies spontaneous symmetry breaking in the boundary theory. This also breaks residual bulk $U(1)$ gauge symmetry generated by $A_{\mu}^3$  spontaneously. The parameters $w_c,\ J_w(=0),\ h_0$ are then specified by using 4 parameters $(a_1,b_1,c_1,z_h)$ of the horizon expansion.

To go further, consider the scaling symmetry in the equations of motion as follows:
\ba\label{TY350}
&(t,y,z)\to \lambda_s^{-1} (t,y,z), \quad \phi\to \lambda_s \phi,\quad w\to \lambda_s w, \\
&f_2\to \lambda_s^2 f_2,\quad h_2\to \lambda_s^{-2} h_2,\quad \phi\to \lambda_s \phi.
\ea
The first symmetry can be used to fix $z_h=1$. Using second symmetry, the leading coefficient of $h_2(z)$ in \eqref{BOU347} can be fixed to set $(h_0=1)$ : this yields the standard $AdS_3$ metric near the boundary.

Under scaling symmetry \eqref{TY350}, the parameters of the solution transform as follows:
\ba
&q\to \lambda_s q,\quad \mu=-q\log (z_p)\to -\lambda_s q\log (z_p)+\lambda_s q\log \lambda_s,\quad \beta \to \lambda_s^{-1} \beta, \nonumber \\
&J_w\to \lambda_s J_w,\quad w_c\to \lambda_s w_c+\lambda_s J_w \log \lambda_s=\lambda_s w_c,
\ea 
where we have used $J_w=0$ becomes zero in the last equality. Note that the scaling transformation of $\log (z)$ in \eqref{BOU347} produces the shift $\log \lambda_s$ { in the chemical potential $\mu$. 

The scaling transformation is important to fix parameters like the temperature or the charge. We can not fix these parameters without changing $z_h$. When we fix the inverse temperature to be $\beta$, we need the scaling transformation $\lambda_s=\beta|_{z_h=1}/\beta$ and then the horizon position $z_h$ is changed into $z_h=\beta /(\beta|_{z_h=1})$ by using the above transformation. To consider the system of the finite charge $q_0$, instead,  we need to perform the scaling transformation $\lambda_s = q_0/q|_{z_h=1}$. The horizon position is then changed into $z_h=q|_{z_h=1}/q_0$.

We find that with a vanishing $J_w$ a solution with non-zero $w_c$ exists only when the charge density $q$ exceeds a critical value $q_c$. This solution is a hairy black hole. As discussed above $w_c$ is the expectation value of a vector order parameter in the boundary field theory : the hairy black hole is then the gravity representation of a superconducting phase.

In figure \ref{fig:ocond} we plot the behavior of  $w_c/q$ as a function of $q_c/q$ at a fixed temperature $T_H=0.15$ with varying $g_{YM}^2$. The critical charge $q_c$ becomes $21.7 T_H$, $33.5 T_H$, and $45.1T_H$  for $g_{YM}^2=5\times 10^{5}$, $10$, and $6$, respectively. The condensate vanishes for $q<q_c$. The point $q=q_c$ is a critical phase transition. Near the critical charge $q_c$, the condensate behaves as $w_c\sim 1.18 q\sqrt{1-q_c/q}$ so that we get a mean field critical exponent.

 The probe approximation corresponds to large $g_{YM}^2$ : this is the situation when the gravity backreaction can be ignored. Our results clearly shows that backreaction 
decreases $w_c/q$ at large charge $q$, while near the critical point the results approach those of the probe approximation. This is expected since near the critical point, $w_c$ is small so that the backreaction is small as well even for finite $g_{YM}^2$.

\begin{figure}[tbph]
  \centering
    \includegraphics[height=4.5cm,clip]{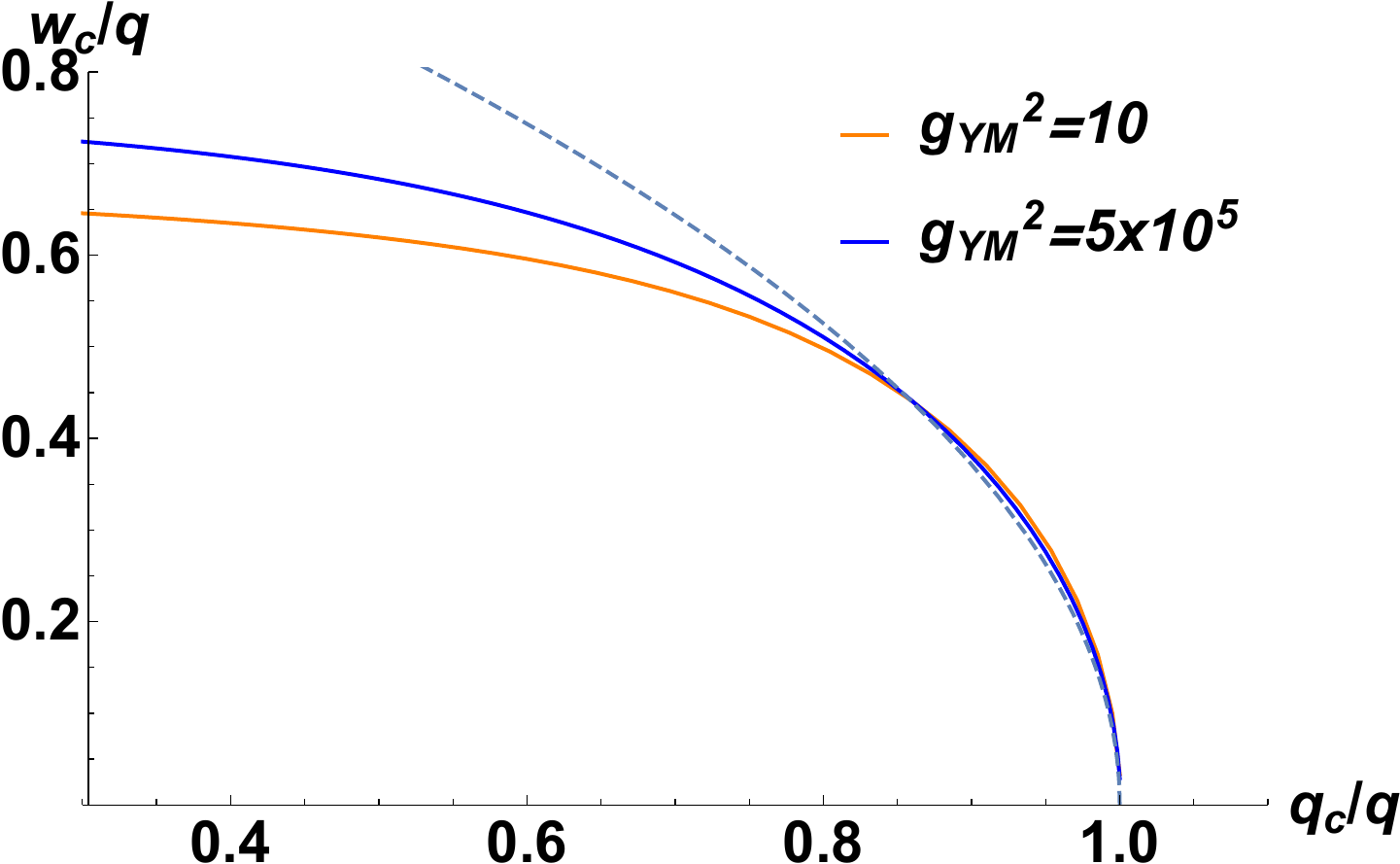} 
  \caption{$w_c$ normalized by $q$ is plotted as a function of $q_c/q$ at the fixed temperature $T_H=0.15$  with varying $g_{YM}^2$. The dashed line shows the analytic curve $1.18\sqrt{1-q_c/q}$. The critical charge $q_c$ is $21.7 T_H$ and $33.5 T_H$  for $g_{YM}^2=5\times 10^{-5}$ and $10$, respectively.}
\label{fig:ocond}
\end{figure}

\begin{figure}[bhtbp]
     \begin{center}
         \includegraphics[height=4.7cm,width=6cm]{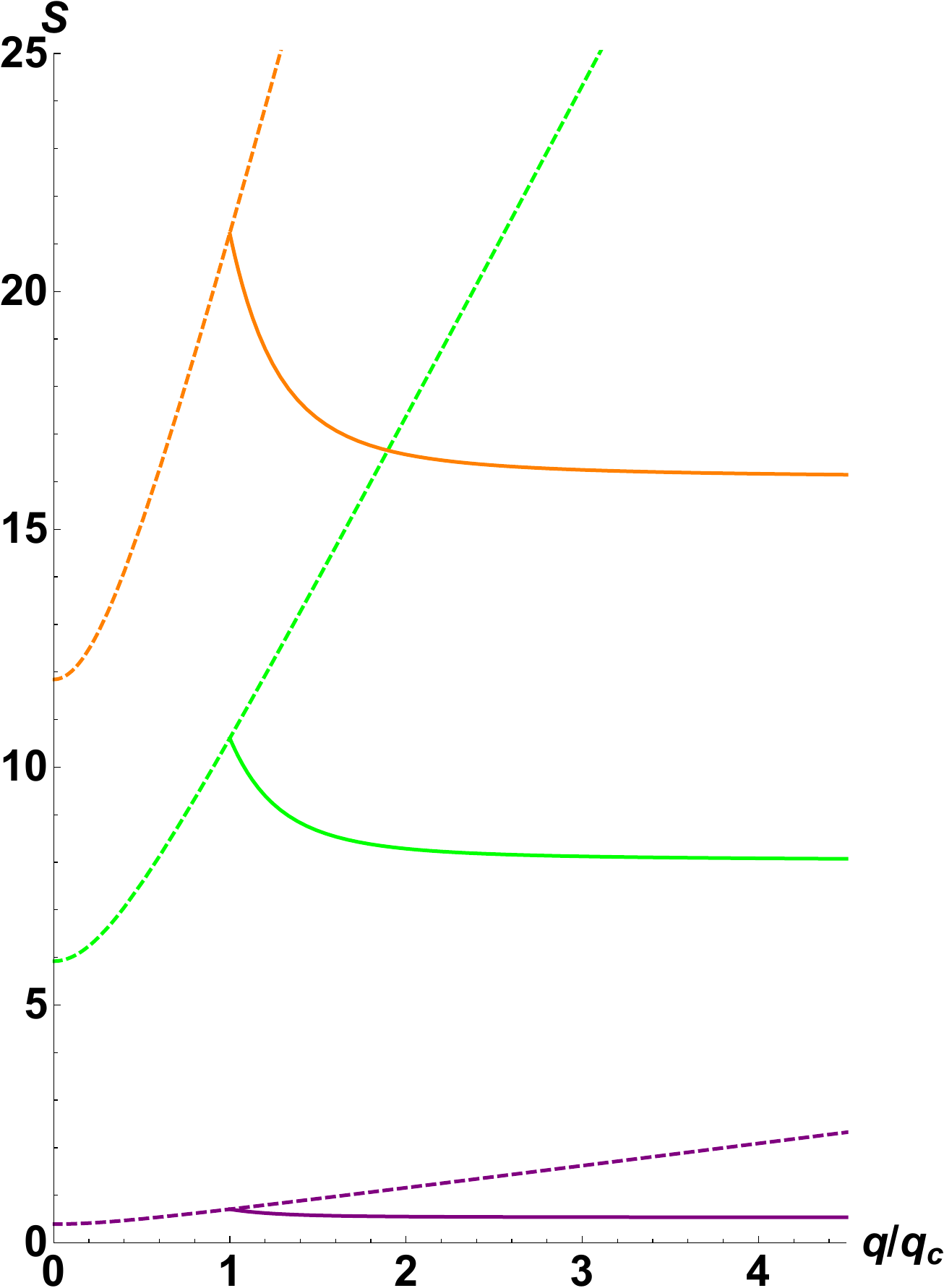} 
         \hspace{1cm}
        \includegraphics[height=4.7cm,width=6cm]{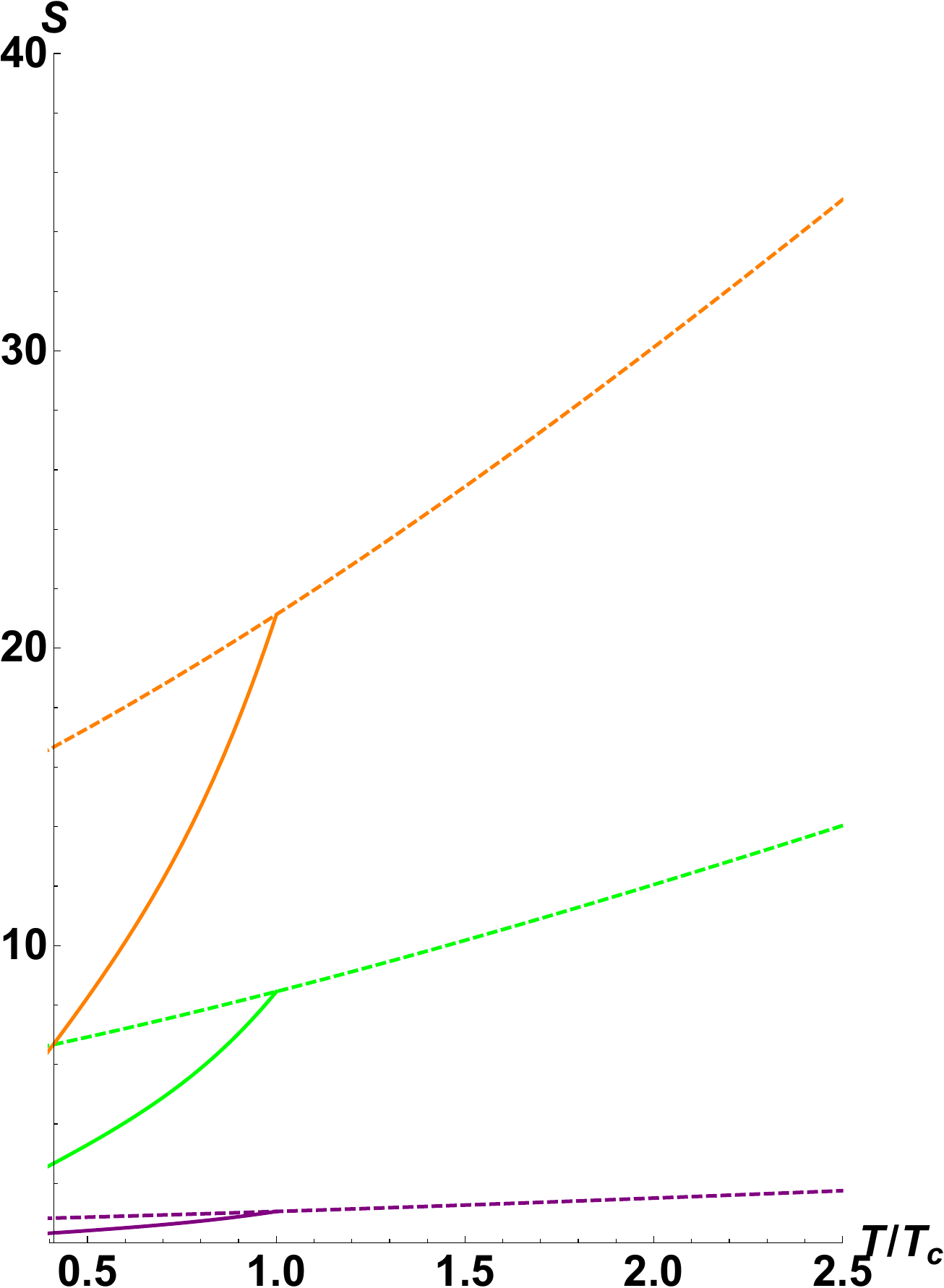} 
           \caption{ The normalized Bekenstein-Hawking entropy $S=2\pi/z_h$ is plotted in units of $2\pi/\kappa^2$ ($= c_{2d}/6$) when $g_{YM}^2=10$. The dashed line gives the entropy in the $AdS_3$ charged black hole. The solid line gives the entropy in the condensed phase $T > T_c$ or $q < q_c$. Left: the normalized BH entropy is plotted for fixed temperature as a function of $q/q_c=q/(33.5T_H)$. Orange, green, and purple curves denote the normalized entropy for $T_H=3/10$, $3/20$, $1/100$, respectively.  Right: The normalized entropy $S=2\pi/z_h$ is plotted for fixed charges as a function of $T_H/T_c$. Orange, green, and purple curves denote the normalized entropy for $q=10$, $4$, and $1/2$, respectively. } 
    \label{fig:entro}
    \end{center}
\end{figure}

In Fig. \ref{fig:entro}, we plot the Bekenstein-Hawking entropy of the hairy black hole $S=(2\pi)^2/(\kappa^2z_h )$ in units of $2\pi/\kappa^2(= c_{2d}/6)$,  where $c_{2d}=12\pi/\kappa^2$ in units of $\tilde{L}=1$ is the central charge of the CFT. 
In the uncondensed phase, the entropy grows like $q^2$ when $q/(g_{YM}T_H) \ll 1$ (the left figure)  and grows like a linear function of $T_H$ when  $q/(g_{YM}T_H)\gg 1$ (the right figure).  We find a cusp at the critical phase transition where $q=q_c(=33.5T_H)$.

\section{Holographic entanglement entropy and the phase transition} \label{sec:two}

In this section we will calculate the entanglement entropy of an interval on the boundary in the $y$ direction of length $l$, using the Ryu-Takayanagi formula~\cite{Ryu:2006bv,Ryu:2006ef,Nishioka:2009un}. The arc on the boundary is chosen to be $-l/2 \leq y \leq l/2$. We then need to calculate the length of a geodesic with minimum length in the bulk metric on a constant time slice which joins the two endpoints of the interval \footnote{For time dependent situations, one needs to consider an extremal geodesic \cite{Hubeny:2007xt}.}. 
\ba 
I_{EE}=\int dy \mathcal{I}_{EE}=\int dy \frac{1}{z} \sqrt{1+\frac{z^{\prime 2}}{h_2(z)f_2(z)}}, \qquad z'=\frac{dz(y)}{dy}.
\ea
Here $z(y)$ denotes a curve in the bulk 2+1 dimensional space-time on a constant time slice. Since we are dealing with a static situation, the action is independent of the time slice.
It is useful to consider the action $I_{EE}$ as the integral of a Lagrangian with $y$ being considered as the time. Since translations of $y$ is a symmetry, the corresponding Hamiltonian is conserved. Thus, the equation of motion which follows from this action becomes a first order differential equation for $z(y)$ :
\ba\label{RADA30}
z'=\sqrt{f_2(z)h_2(z)\Big(\frac{z_*^2}{z^2}-1\Big)} ,
\ea
where $z=z_*$ is the turning point, i.e. the point where $z'(y)$ vanishes. The curve $z(y)$ is assumed to be smooth everywhere. 

The length $l$ of the interval can be now easily calculated
\ba
l_{curve}=2\int^{\infty}_{z_*}dz\dfrac{1}{\sqrt{f_2(z)h_2(z)\Big(\frac{z_*^2}{z^2}-1\Big)}}.
\ea
We have assumed that the curve is symmetric under reflections about the turning point - this gives the factor of 2 in front of $l_{curve}$
A solution to the equation (\ref{RADA30}) with specified end points is a minimal length geodesic joining those points.
According to the Ryu-Takayanagi formula the holographic entanglement entropy is the on-shell action of $I_{EE}$ divided by the gravitational constant
\ba\label{ACTA31}
\see=\dfrac{2\pi}{\kappa^2}I_{EE}|_{\text{on-shell}}=\dfrac{4\pi}{\kappa^2}\int^{\epsilon}_{z_*} dz\dfrac{1}{z\sqrt{f_2(z)h_2(z)\Big(1-\frac{z^2}{z_*^2}\Big)}}. 
\ea
One can check that $l\to \lambda_s^{-1}l$ under the rescaling $(z,z_h,z_*,\epsilon)\to \lambda_s^{-1}(z,z_h,z_*,\epsilon)$. It is then convenient to introduce the scale invariant quantities $l q$ or $l T$. 
Note that the holographic entanglement entropy is invariant under the scaling transformation varying $(z,z_h,z_*,\epsilon)$. In equation (\ref{ACTA31}) the boundary $z=0$ has been replaced by a cutoff boundary $z = \epsilon$ (which is essentially a UV cutoff of the boundary field theory). This regulates the UV divergence of the entanglement entropy, as discussed below.

We are interested in calculating $\see$ as a function of the charge $q$ for a given temperature. As discussed above, for $q < q_c$ the relevant bulk geometry is a charged black hole, while for $q > q_c$ the background is given by a hairy black hole with a nonzero condensate $w_c$, with the metric \eqref{ANS13} in section 2.
In both cases, the UV divergent (area law) term of $I_{EE}$ behaves like $I_{EE}\sim -2\log (\epsilon)$.  This motivates the definition of the finite part of the minimal length,
\be
I_{m,fin}\equiv I_{EE}+2\log (\epsilon)
\label{finite}
\ee
$I_{m,fin}$ is independent of the cutoff. Under the scaling transformation discussed above, $I_{m,fin}$ changes only by the additive constant $-2\log (\lambda_s)$, which will not affect our analysis qualitatively. 

We now use the metric obtained in the previous section for both the condensed and uncondensed phase to calculate $I_{m,fin}$. This involves evaluation of the integral (\ref{ACTA31}), which we perform numerically.

In the uncondensed phase $T>T_c$  or $q<q_c$, we find that the holographic entanglement entropy is proportional to the subsystem size $l$ ( a volume law ) when {the length $l$ is large ($lT_H \gg1$) exactly as in higher dimensions~\cite{Tonni:2010pv,Kundu:2016dyk}. In the opposite limit $T_Hl\ll 1$ and $ql\ll 1$, $I_{m,fin}$ approaches the value in pure $AdS_3$, namely, $\see \sim 4\pi/\kappa^2\cdot \log (l/a)=c/3\cdot \log (l/a)$. These behaviors in the extreme limits are of course what is expected. Interesting non-trivial behavior is expected for $T$ or $q$ in the intermediate range~\cite{Belin:2013uta}, particularly when they are close to their values at the critical superconducting transition.

\begin{figure}[t!]
     \begin{center}
     \includegraphics[height=5.5cm,clip]{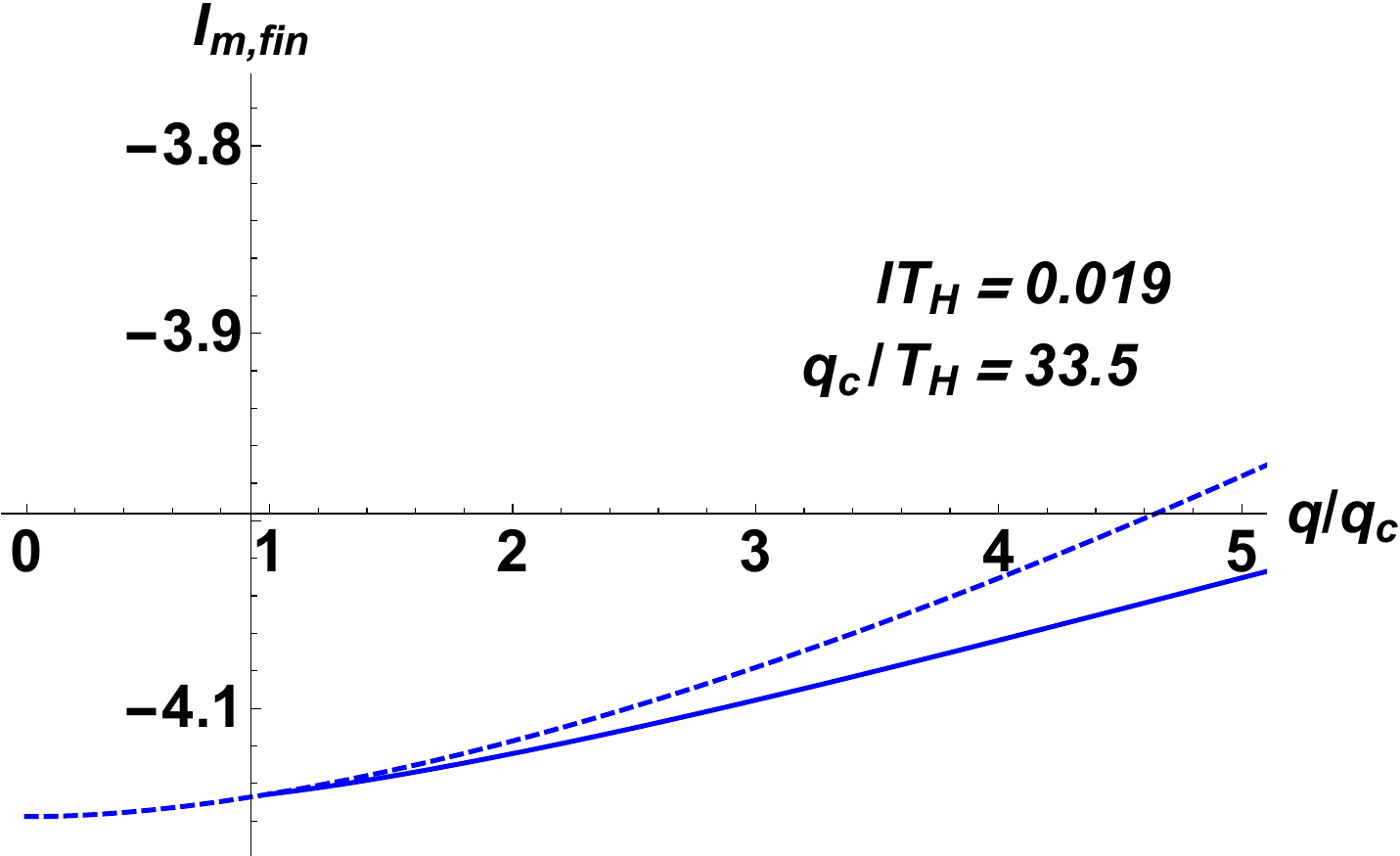} 
     \caption{$I_{m,fin}$ as a function of $q/q_c$ at $lT_H =0.019$ with $T_H=0.15$ and $g_{YM}^2=10$. The dashed curve is for a $AdS_3$ charged black hole corresponding to the uncondensed phase, $q < q_c$. The solid curve is for the condensed phase described by a hairy black hole. } 
    \label{fig:cond}
    \end{center}
\end{figure}

Figure (\ref{fig:cond}) shows the finite part of the entanglement entropy $I_{m,fin}$ for an interval of length $l$ as a function of $q/q_c$ for $g_{YM}^2=10$ for a dimensionless length $lT_H =0.019$. The dashed curve is $I_{m,fin}$ in the uncondensed phase described by the $AdS_3$ charged black hole. The solid curves is the result in the condensed phase. In both phases the finite part of the entanglement entropy increases with increasing charge. However there is a cusp at the critical point $q=q_c$ where $\frac{dI(l,fin)}{dq}$ is discontinuous. Note that in the condensed phase $I_{m,fin}$ is always smaller than the value of this quantity in the background of a charged black hole. This is consistent with the fact that condensation results in a depletion of the number of degrees of freedom.

\begin{figure}[t!]    
   \begin{center}
        \includegraphics[height=4.5cm,clip]{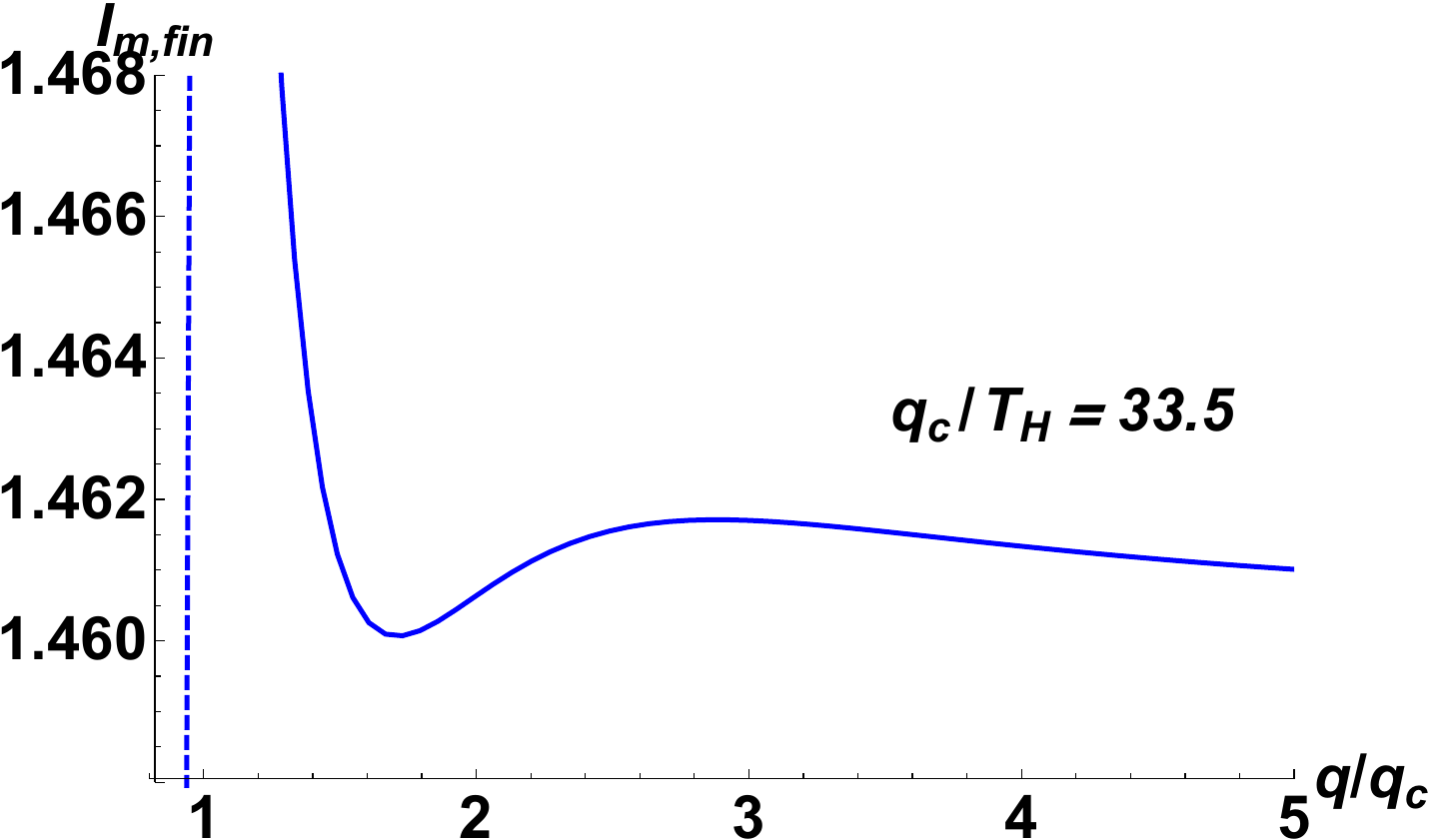} 
        \includegraphics[height=4.5cm,clip]{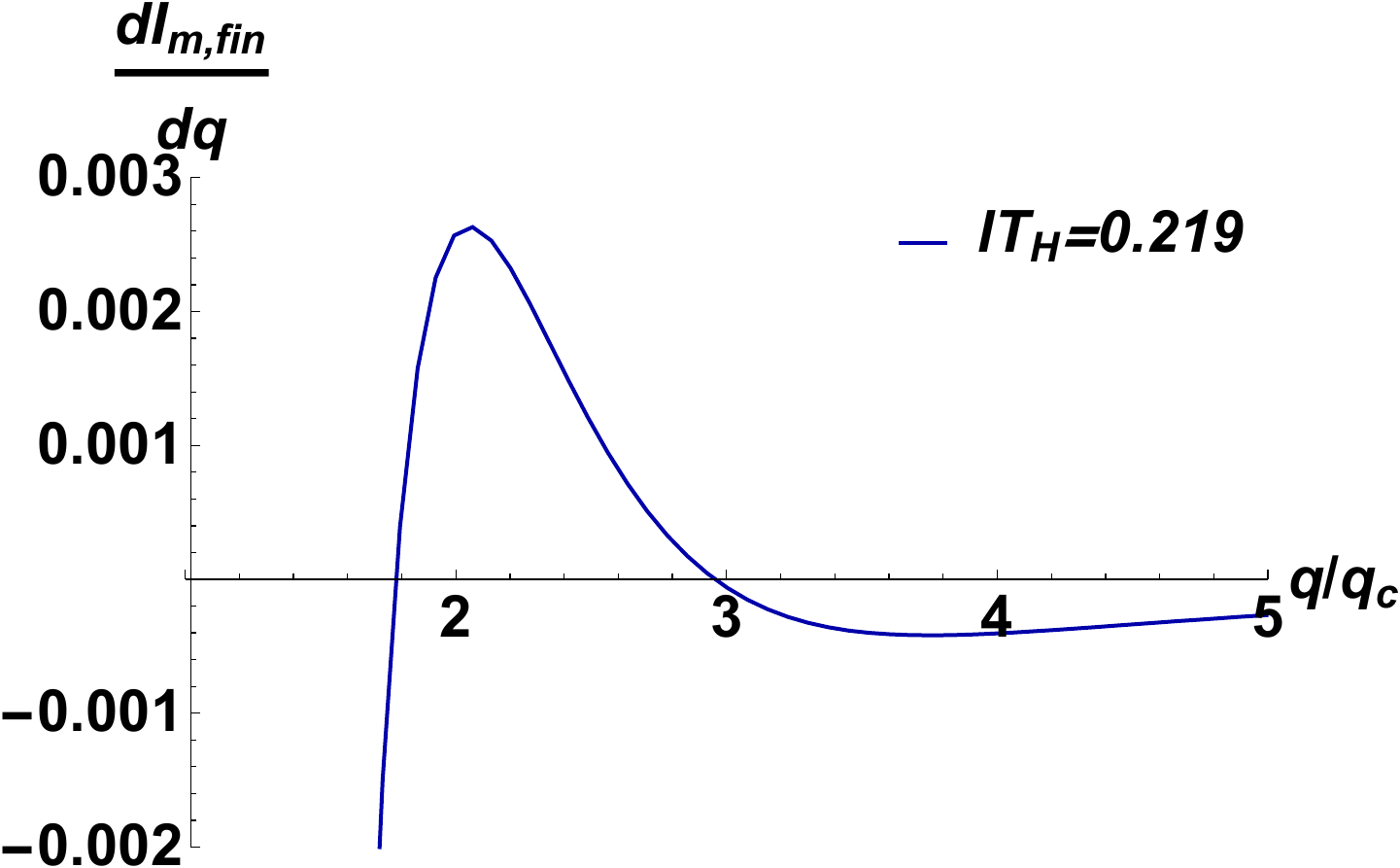}
\caption{Typical non-monotonic behavior of $I_{m,fin}$ as a function of $q/q_c$ for $lT_H=0.219$ for  $g_{YM}^2=10$ depicted in the left plot. Right: Its derivative with respect to charge. One can see clearly two zeros: the first one is minimum at $q=1.8 q_c$, while the second one is maximum at $q=3 q_c$. } 
    \label{fig:cond2}
    \end{center}
\end{figure} 

The monotonic increase of $I_{m,fin}$ with increasing charge continues to larger interval lengths till we reach a critical interval size $l_{c1}$. For $g_{YM}^2 = 10$ we have $l_{c1}T_H \sim 0.15$. For $l > l_{c1}$  there is a non-monotonic behavior : typically, $I_{m,fin}$ first decreases beyond the cusp at the critical point, reaches a minimum and increases to reach a maximum, and then approaches a plateau.  
This kind of non-monotonic behavior is shown in figure (\ref{fig:cond2}) for an interval size $lT_H = 0.219$. There are several notable aspects of the behavior. First, as we increase $lT_H$ the minimum of $I_{m,fin}$ is pushed to values of $q$ away from the critical point $q=q_c$. In fact the critical length $l_{c1}$ is the value of $l$ where this local minimum is exactly at the critical point. Secondly, $I_{m,fin}$ approaches a plateau at large $q$. The value of $I_{m,fin}$ at the plateau is larger than the value of $I_{m,fin}$ at $q=q_c$ for small interval lengths (still larger than $l_{c1}$) whereas at larger interval lengths the value of $I_{m,fin}$ at the plateau is {\em smaller} than the value of $I_{m,fin}$ at $q=q_c$. Finally, the difference of the value of $I_{m,fin}$ and its minimum value decreases as we increase $l$.

Consequently, there is another critical interval size $l_{c2}$. For $l > l_{c2}$ the quantity $I_{m,fin}$ decreases monotonically for $q > q_c$. For $g_{YM}^2 = 10$ we get $l_{c2}T_H \sim 0.22$. At this value of $l$ the minimum of $I_{m,fin}$ is pushed beyond the largest value of $q$ for which we could do the numerics, while the plateau value becomes very close to the minimum value. 

This complex behavior is possibly a result of two competing trends. First, an increasing charge density tends to increase the entanglement entropy. This is clear in the uncondensed phase for all values of $l$. However condensation beyond $q = q_c$ results in a depletion of the number of degrees of freedom, and therefore tend to decrease the entanglement entropy. For small enough interval sizes, the effect of condensation is not very pronounced, so that the first trend wins leading to a monotonically increasing $\see$. In the dual gravity picture the condensate is pronounced near the horizon - the minimal geodesic stays far from the horizon for small intervals. This is consistent with the fact that at length scales which are significantly smaller than the scale of symmetry breaking, effects of symmetry breaking are invisible. However as the interval size increases, the minimal geodesic goes deep into the bulk and the effect of the condensate becomes pronounced. The intermediate region $ l_{c1} \leq l \leq l_{c2}$ is possibly characterized by the regime where these two effects are of the same order and therefore compete. Finally, for $l > l_{c2}$ the effect of condensation overcomes the effect of increasing charge density. The degrees of freedom keep decreasing as we increase the charge density resulting in a monotonic decrease of the $\see$.

While we not have a quantitative calculation to back the above scenario we can check its consistency by calculating $I_{l,fin}$ for different values of $g_{YM}^2$ and therefore changing the strength of backreaction. As we will see, this turns out to give further insights into the nature of the critical values of the critical sizes of $l_{c1}$ and $l_{c2}$ and the mechanism of the competition. We present the $g^2_{YM}=10$ case first, followed by $g^2_{YM}=6$ and $g^2_{YM}=50$. The smaller the value, the stronger the back-reaction and the effects of the competition mentioned above become more pronounced.

The entanglement entropy $\see$ , normalized as $ I_{m, fin}(I,q)/I_{m, fin}(I,q_c)$, for $g^2_{YM}=10$ with 6 different sub-system sizes is presented in Fig. \ref{fig:gym10}, so that all the $\see$ coincide at the critical point $q=q_c$. In the left figure of Fig. \ref{fig:gym10}, we show the behavior of entanglement entropy as we increase the sub-system size from the top plot to the bottom plot. The right figure is the slope of the entanglement entropy with respect to the charge, $ dI_{m, fin}(I,q)/dq$. In the range $q>q_c$, the slope is positive for $lT_H=0.14$ and thus entanglement entropy increases monotonically, while the slope is negative for $lT_H=0.23$ with monotonically decreasing entanglement entropy. The entanglement entropy for $0.21 \lesssim lT_H \lesssim 0.22$ have non-monotonic behaviors revealing two zeros of its derivative. This is shown in the right panel of Fig. \ref{fig:gym10closeup}. As we increase the charge, entanglement entropy decreases, hits a minimum (first zero of the derivative), increases, arrives at a maximum (second zero of the derivative), and eventually reaches a plateau. Typically, the second zero corresponding to a local maximum happens at a very large value of $q/q_c$.
Finally, we clearly see that the minimum of the entanglement entropy moves to $q=q_c$ for $lT_H=0.15$, while for $lT_H=0.22 $ the minimum and the maximum of the entanglement entropy coincide, forming a point of inflection.  

\begin{figure}[t!]
	\begin{center}
		\includegraphics[height=5cm,clip]{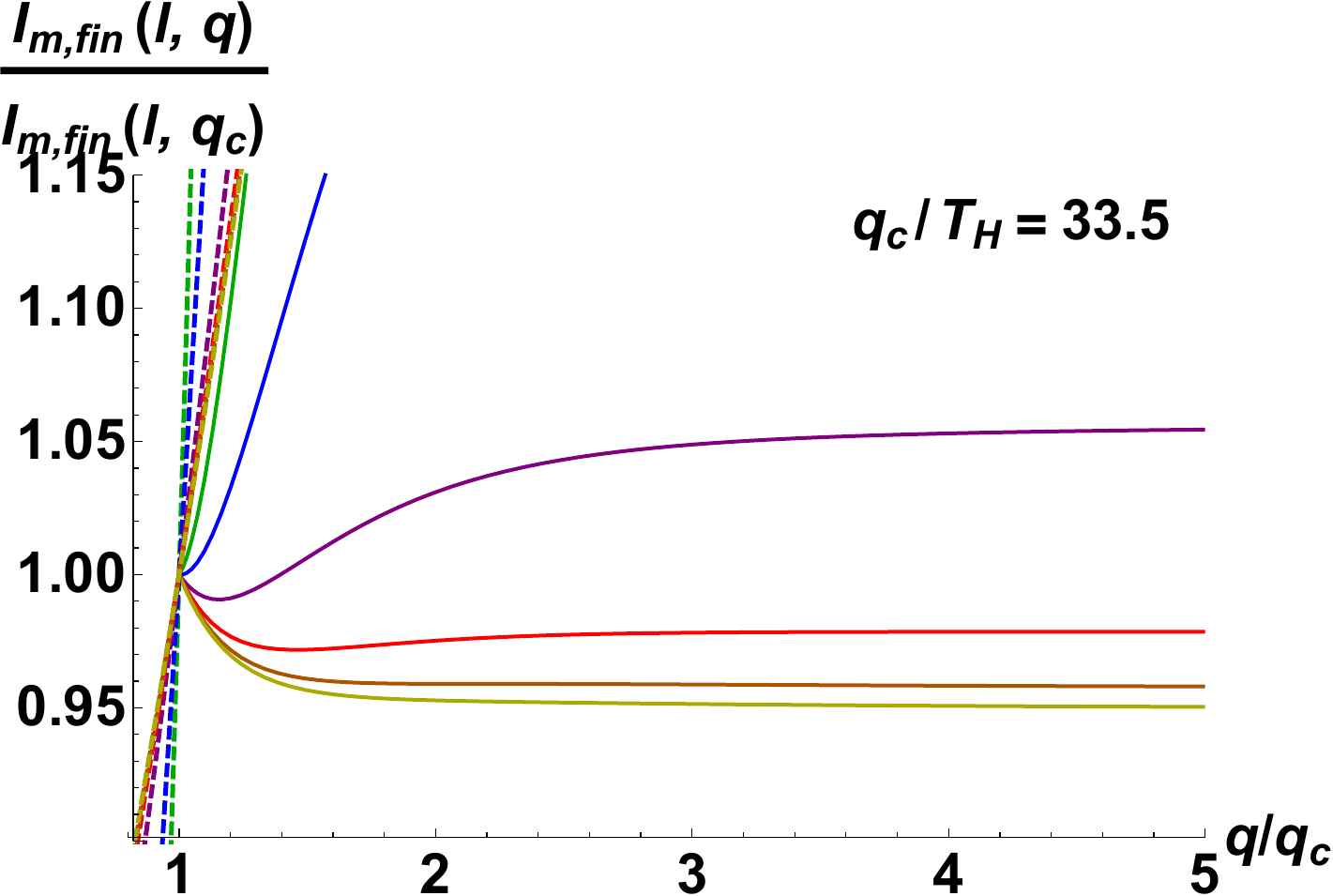} 
		\hspace{0.2cm}
		\includegraphics[height=5cm,clip]{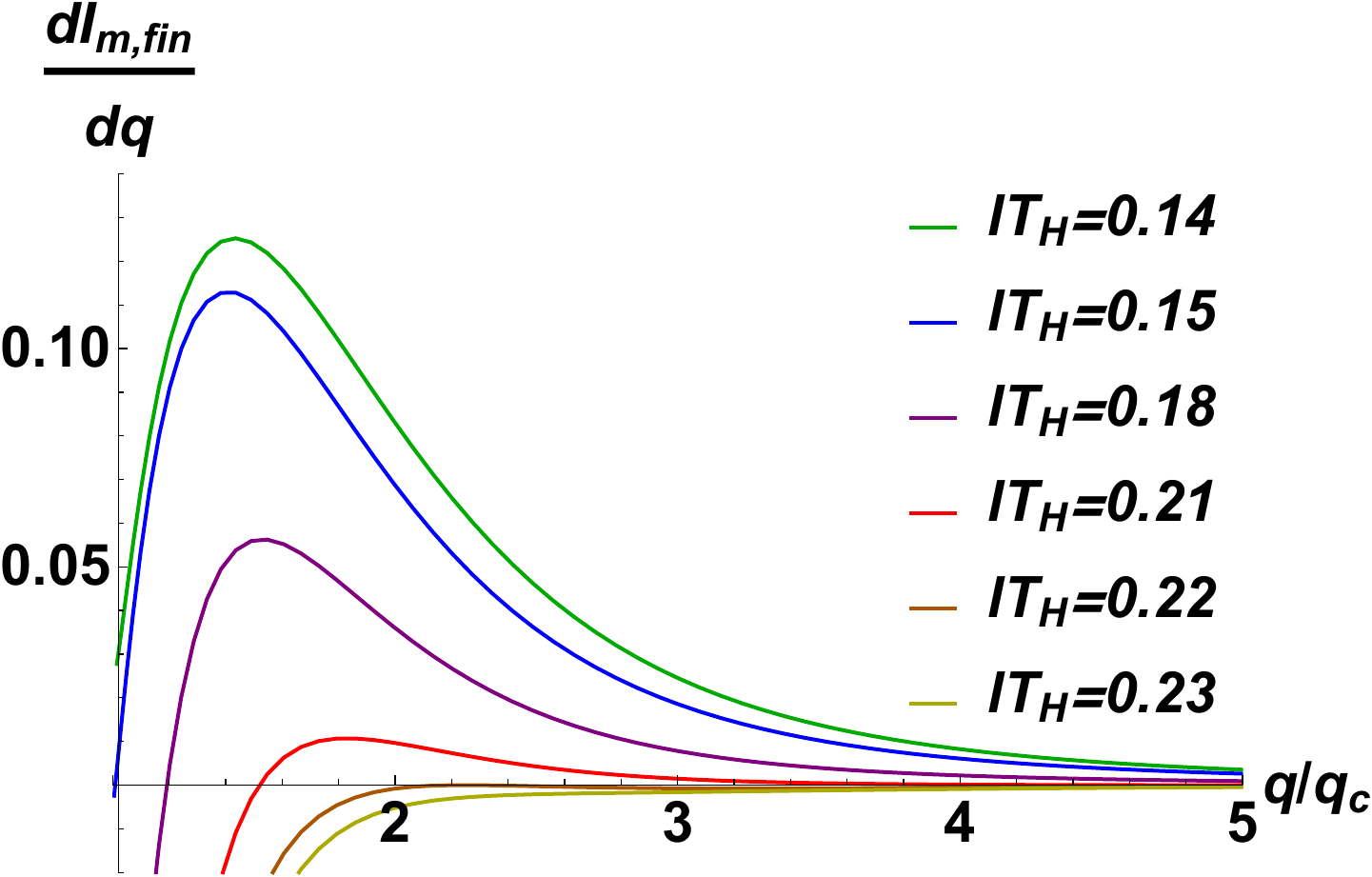} 
		\caption{Normalized entanglement entropy as a function of $q/q_c$ for 6 different invariant sub-system sizes $lT_H$ (from top to bottom, $lT_H=0.14,\ 0.15,\ 0.18,\ 0.21,\ 0.22,\ 0.23$) with a fixed temperature $T_H=0.15$ and $g_{YM}^2=10$. The dashed and solid curves are for uncondensed and condensed phases, respectively.} 
		\label{fig:gym10}
	\end{center}
	\begin{center}
		\includegraphics[height=4.5cm,clip]{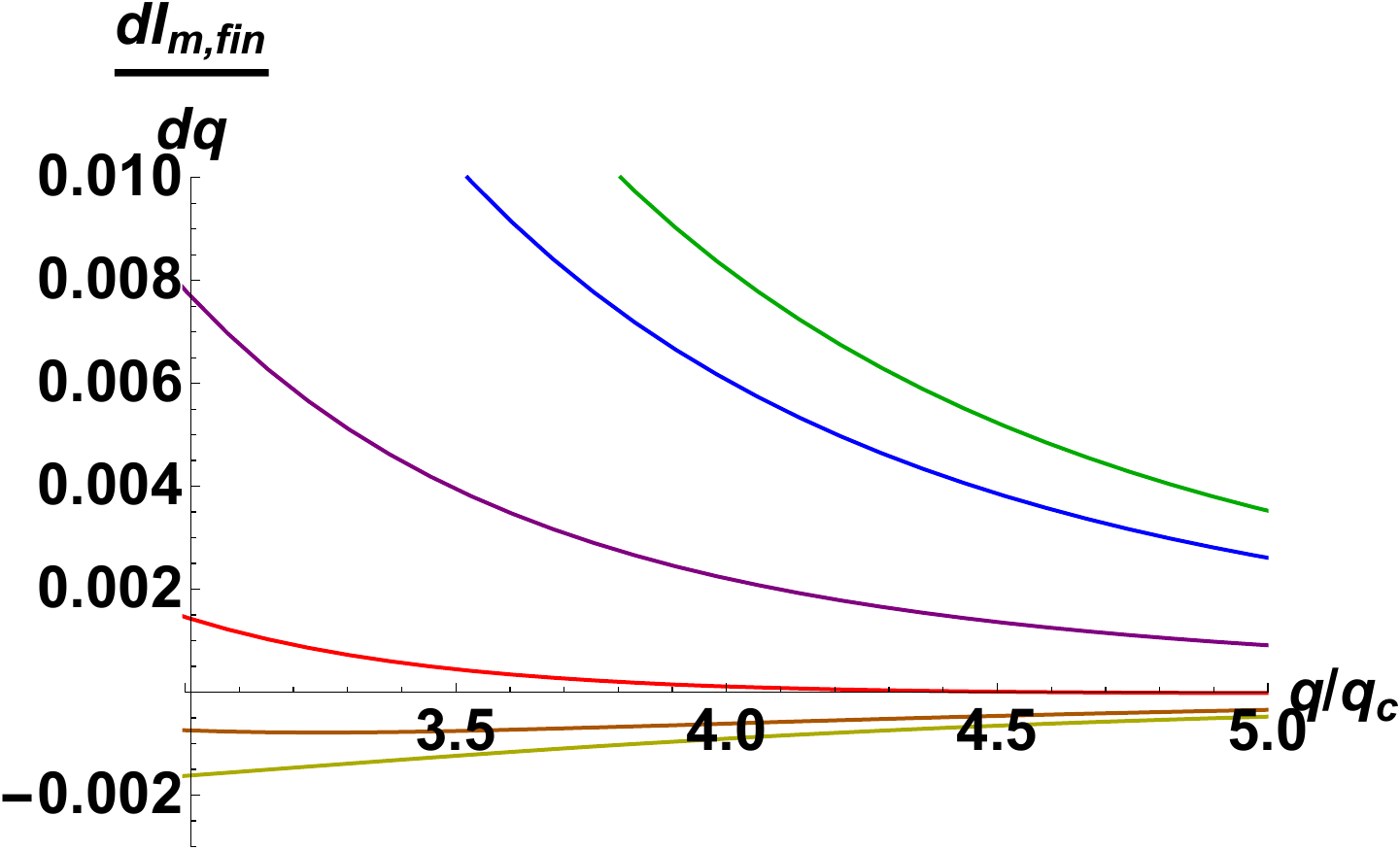} 
		\hspace{0.2cm}
		\includegraphics[height=4.5cm,clip]{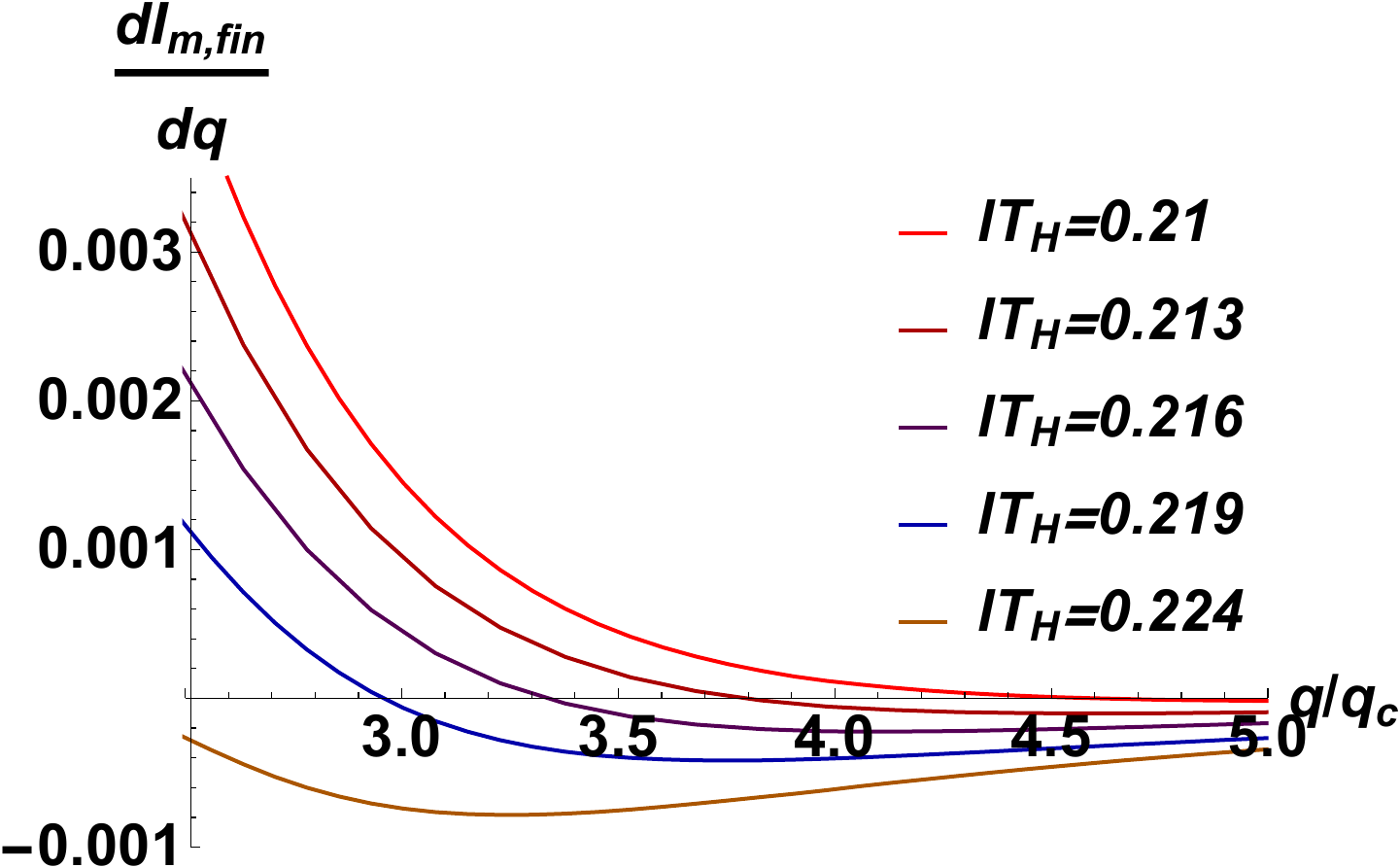} 
		\caption{Left: A plot enlarging the region $3\le q/q_c\le 5$ of Fig. \ref{fig:gym10}. We have used the same color for different invariant sub-system sizes $lT_H$. Right: More details of the second zeros by changing $lT_H$ slightly.} 
		\label{fig:gym10closeup}
	\end{center}
\end{figure}

\begin{figure}[t!]
     \begin{center}
        \includegraphics[height=5.5cm,clip]{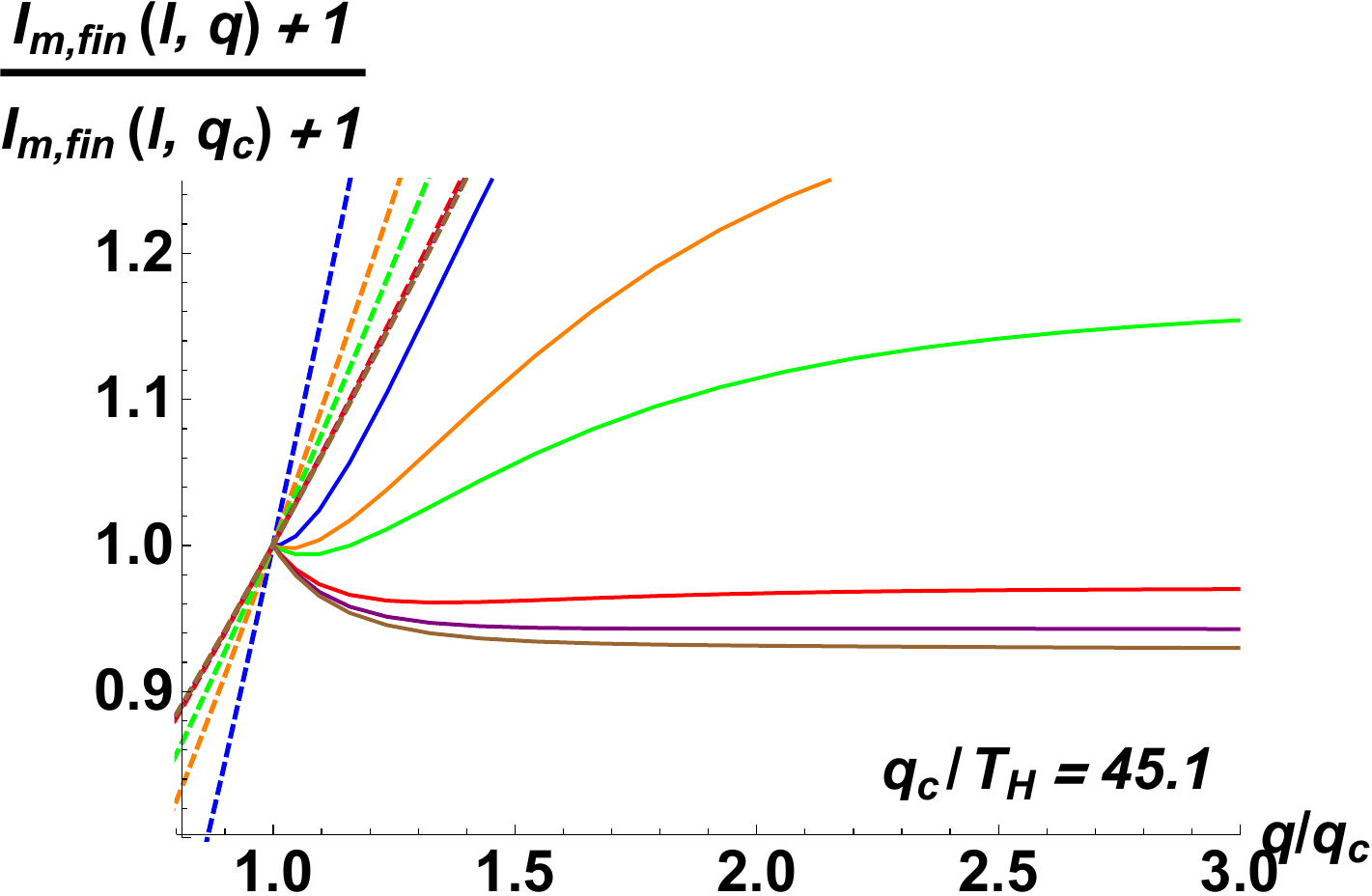} 
           \hspace{0.1cm}
          \includegraphics[height=6.5cm,clip]{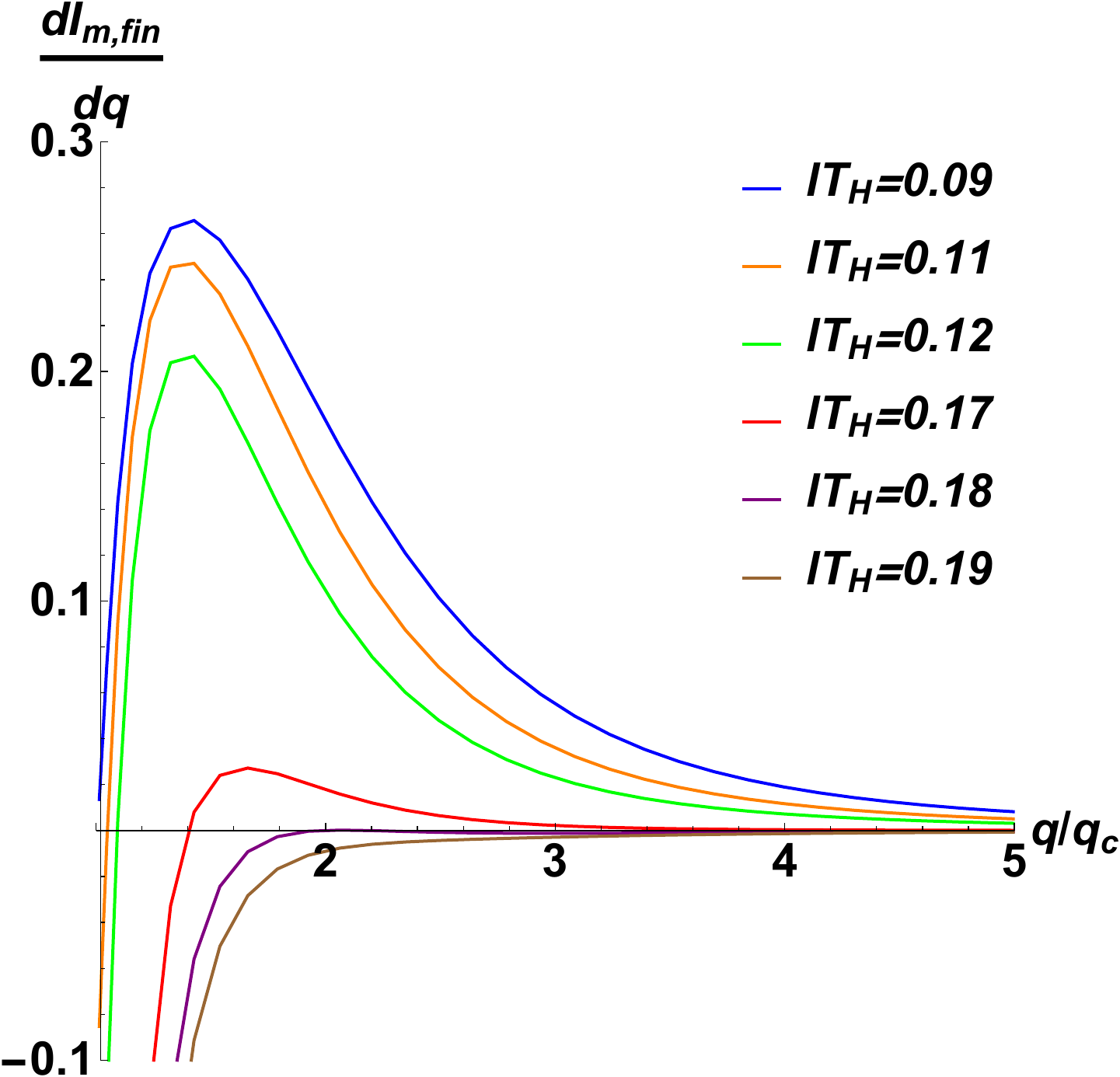} 
           \caption{Normalized entanglement entropy as a function of $q/q_c$ for $g_{YM}^2=6$. We plot the entanglement entropy for 6 different invariant sub-system sizes $lT_H$  (from top to bottom, $lT_H=0.09,\ 0.11,\ 0.12,\ 0.17,\ 0.18,\ 0.19$) with a fixed temperature $T_H=0.15$. The dashed and solid curves are for uncondensed and condensed phases, respectively.} 
    \label{fig:gym6}
    \end{center}
\end{figure}

\begin{figure}[t!]
	\begin{center}
		\includegraphics[height=5cm,clip]{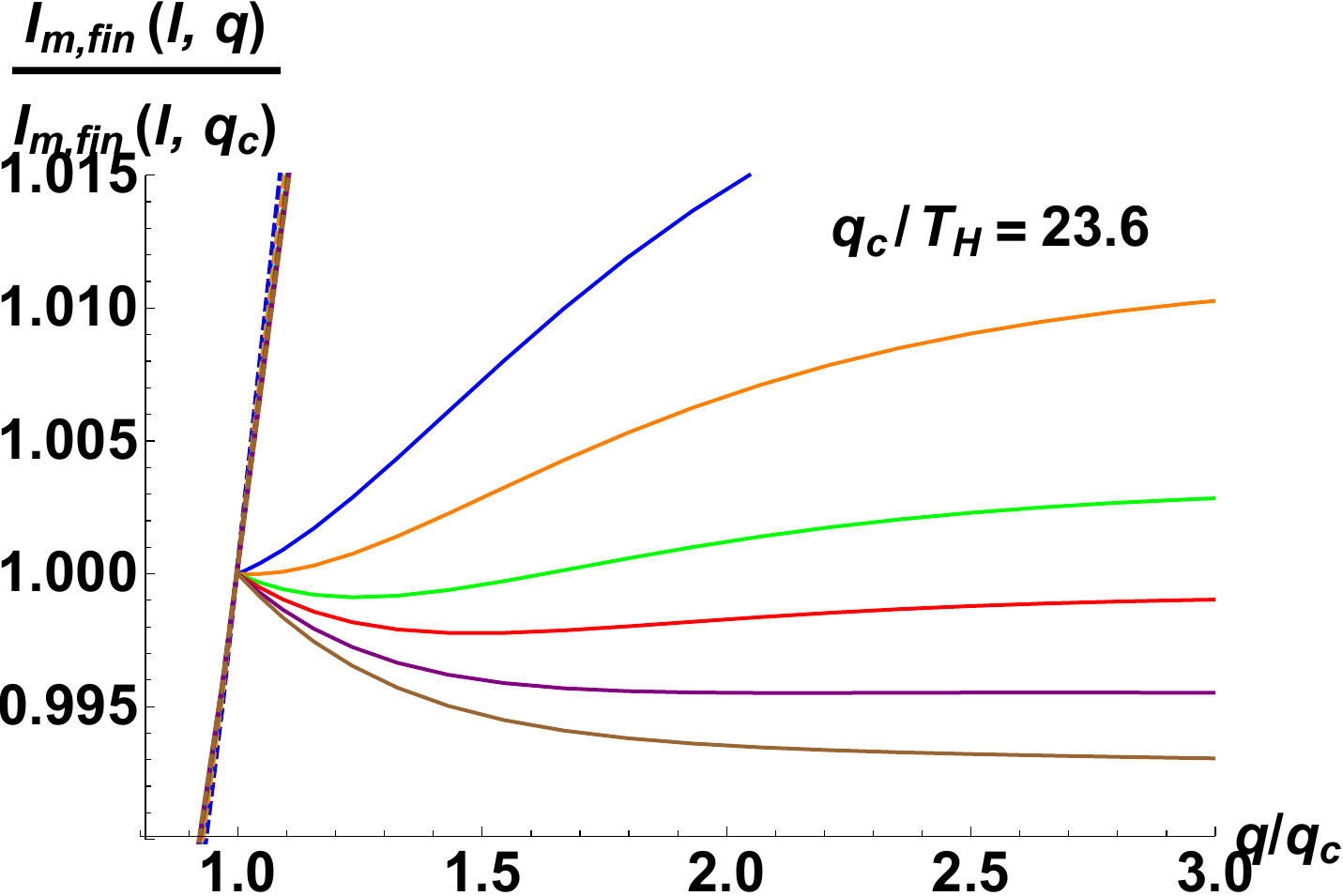} 
		\hspace{0.1cm}
		\includegraphics[height=5cm,clip]{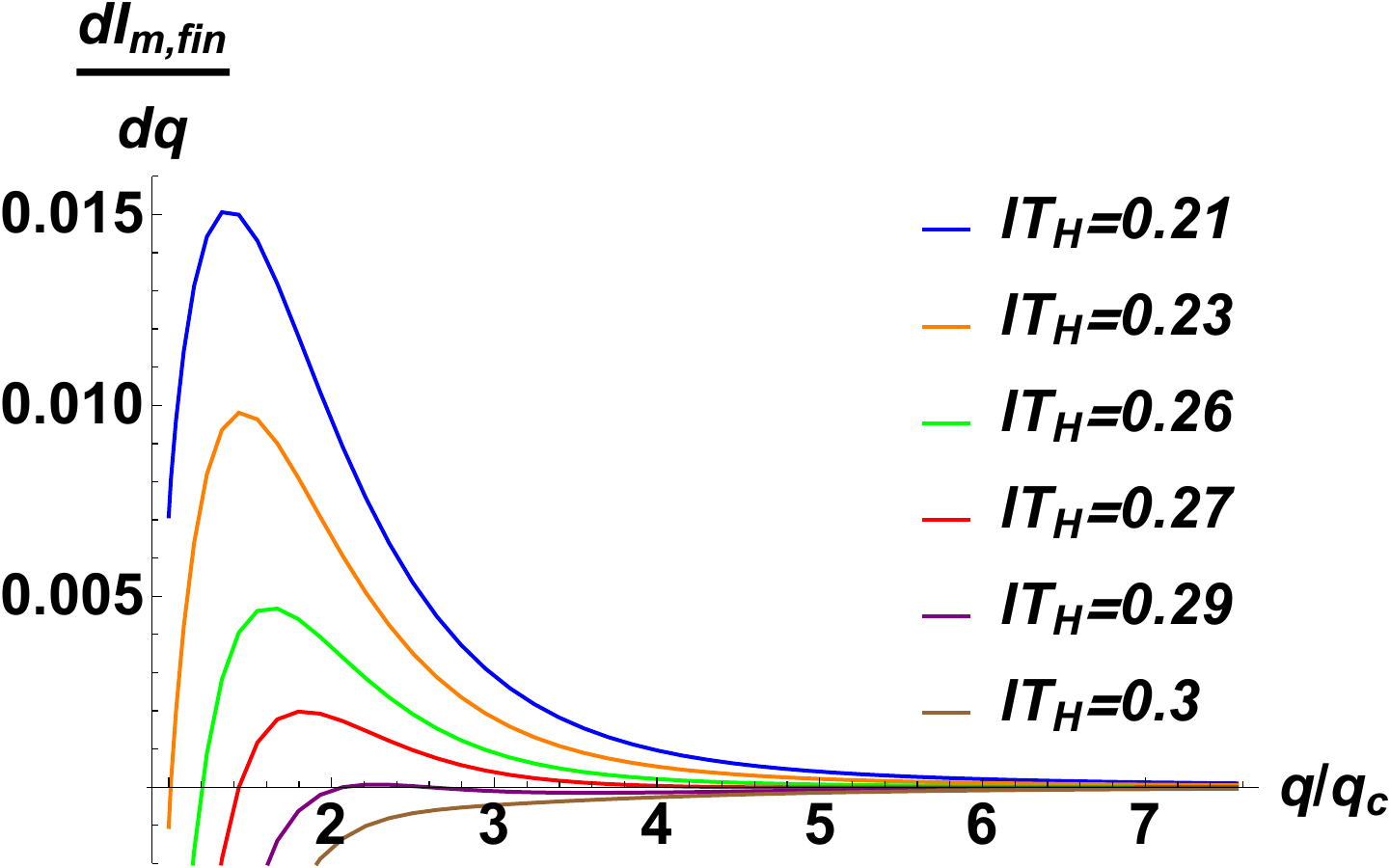} 
		\caption{Normalized entanglement entropy as a function of $q/q_c$ for $g_{YM}^2=50$. We plot the entanglement entropy for 6 different invariant sub-system sizes $lT_H$  (from top to bottom, $lT_H=0.21,\ 0.23,\ 0.26,\ 0.27,\ 0.29,\ 0.3$) with a fixed temperature $T_H=0.15$. The dashed and solid curves are for uncondensed and condensed phases, respectively. } 
		\label{fig:gym50}
	\end{center}
\end{figure}

Similarly, we present entanglement entropy for $g^2_{YM}=6$ in Fig. \ref{fig:gym6} and for $g^2_{YM}=50$ in Fig. \ref{fig:gym50} with  different sub-system sizes. The entnaglement entropy shows behaviors similar to those at $g^2_{YM}=10$. For $g^2_{YM}=6$, non-monotonic behavior can be observed for $0.09 \lesssim l T_H \lesssim 0.18$ with two critical points located at $l_{c1}T_H=0.09$ and $l_{c2} T_H= 0.18$. For $g^2_{YM}=50$, non-monotonic behavior exists for $0.22 \lesssim l T_H \lesssim 0.29$ that is sandwiched by two critical points $l_{c1}T_H=0.22$ and $l_{c2} T_H= 0.29$.   
Combining together we see the following pattern. 
\[ \left. \begin{array}{c|cc|c}
 & ~~l_{c1}T_H~~ & ~~l_{c2} T_H~~ & ~~\Delta l T_H = l_{c2}T_H - l_{c1}T_H~~\\ \hline
g^2_{YM} =6 & 0.09 & 0.18 & 0.09 \\
g^2_{YM} =10 & 0.15 & 0.23 & 0.08 \\
g^2_{YM} =50 & 0.22 & 0.29 & 0.07  \\
g^2_{YM} =250 & 0.25  & 0.30  & 0.05  \end{array} \right.\] 
As we increase the strength of the back reaction, decreasing $g^2_{YM}$, both critical lengths   $l_{c1}T_H$ and $l_{c2}T_H$ decrease while the invariant distance between them, $\Delta l T_H = l_{c2}T_H - l_{c1}T_H$, increases.

 \subsection{A Phase Diagram} 
 
As discussed above, the non-monotonic behavior of the entanglement entropy is a result of a competition between two effects : the effect of charge density and the effect of condensation. It is useful to re-cast this in terms of two length scales which emerge in the problem.

Let us recapitulate the salient features of our results in Fig 3 - Fig 8. These are all plots of the entanglement entropy as a function of the charge density, the different curves being results for different values of $(lT_H)$. For a given value of $g_{YM}$ we have the following :

\begin{itemize}

\item{} For $l > l_{c2}$ the entanglement entropy decreases monotonically as a function of $q/q_c$

\item{} At $l = l_{c2}$ there is a point of inflextion at $q = q^\star$

\item{} For $l_{c2} > l > l_{c1}$ the entanglement entropy has a minimum at some value $q_{min} (l)$ and a maximum at a larger value $q_{max} (l)$. 

\item{} The minimum $q_{min} (l)$ {\em increases} as a function of $l$. At $l = l_{c1}$ one has $q_{min} (l_{c1}) = q_c$. On the other hand, $q_{max} (l)$ {\em decreases} with increasing $l$, reaching its minimum value $q^\star$ at $l = l_{c2}$. Thus at $l=l_{c2}$ the maximum and the minimum merge into a point of inflection.

\item{} At $l=l_{c1}$ the minimum is at $q=q_c$

\item{} For $l < l_{c1}$  the entanglement entropy increases as a function of $q$ at least upto $q \sim5 q_c$ which is the maximum value of $q$ used in our calculation. 

\end{itemize}

It is then clear that the minimum of the entanglement entropy as a function of the charge can occur only for $q < q^\star$, while the maximum of the entanglement entropy as a function of the charge can occur only for $q > q^\star$.

To rephrase the behavior in terms of length scales it is useful to plot the quantity $(\partial I_{m,fin}/ \partial q)_l$ as a function of $lT_H$. This is shown in Fig (\ref{fig:LengthScale22}). 
 
 \begin{figure}[t!]
 	\begin{center}
 		\includegraphics[height=5cm,clip]{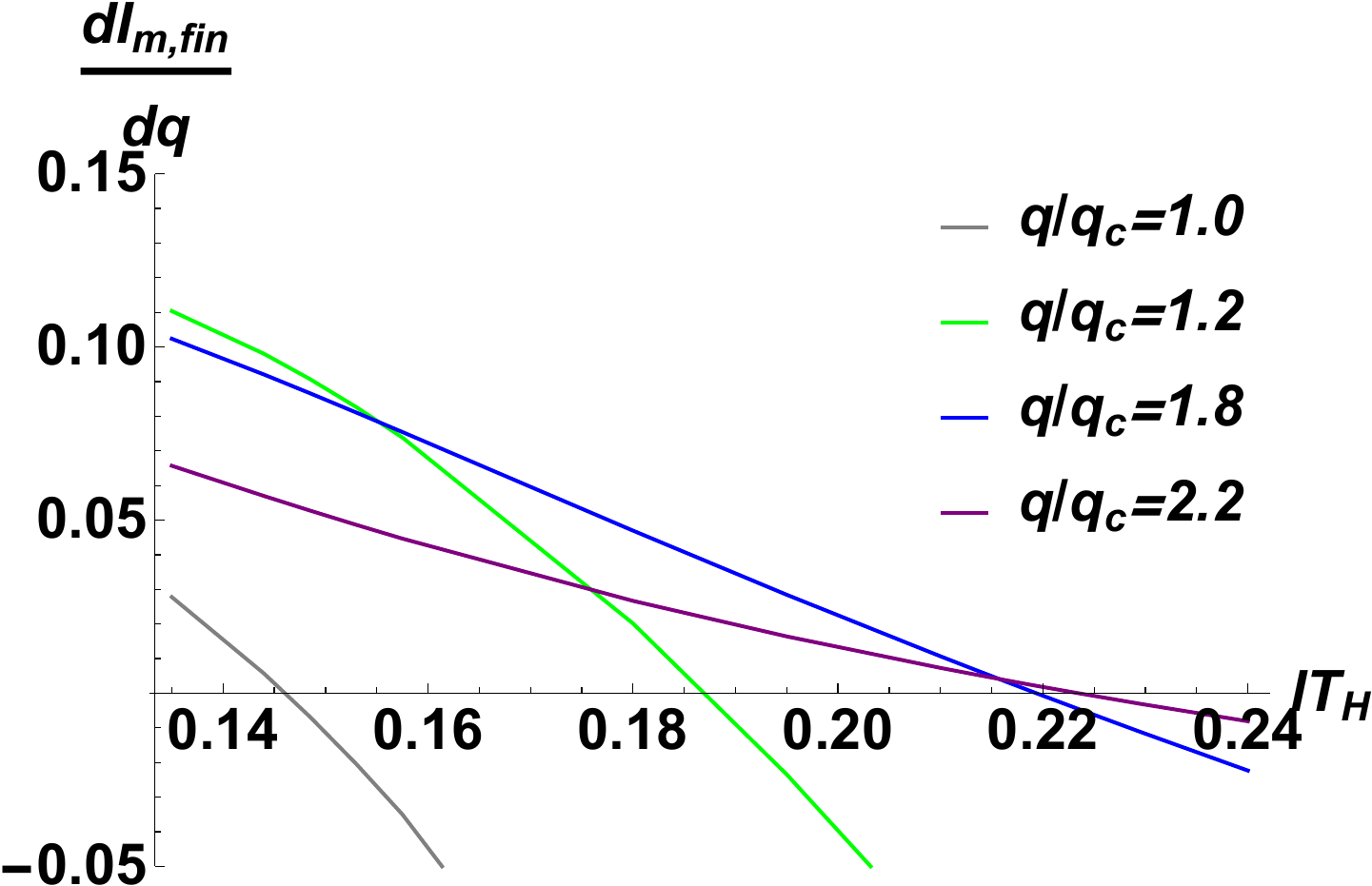} 	\includegraphics[height=5cm,clip]{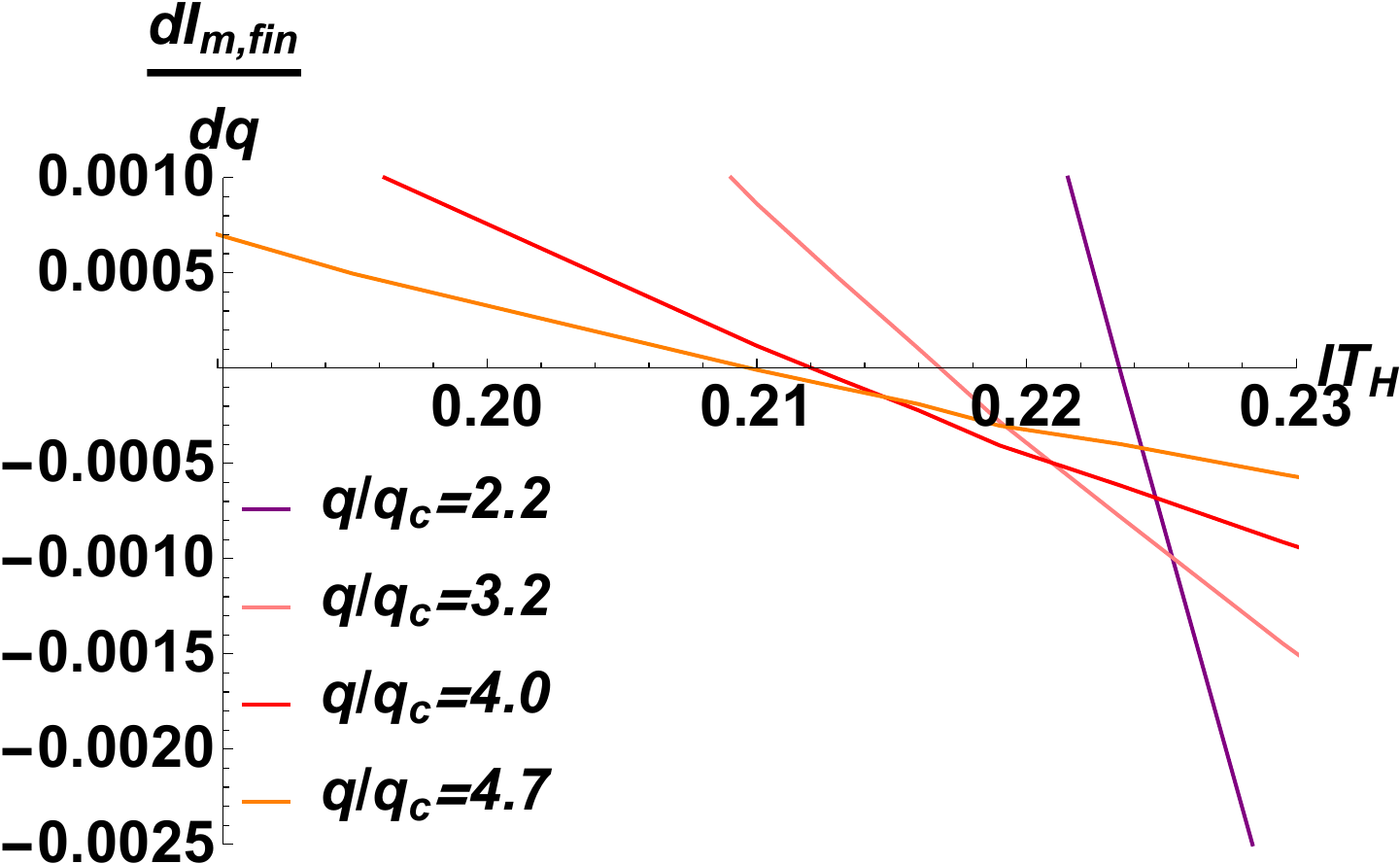}
 		\caption{\small $\partial I_{m,fin}/\partial q$ as a function of the sub-system size $l$ for $g_{YM}=10$ and for various values of $q/q_c$. The left panel shows the results for $q < q^\star$, where $q^\star$ is in between $q/q_c=2 $ and $q/q_c=3$. The zero of $\partial I_{m,fin}/\partial q$ in this regime of charge is a minimum (as a function of $q$), and the length scale $L_c$ is defined as the value of $l$ at which $\partial I_{m,fin}/\partial q = 0$. Clearly, $L_c$ increases as the charge density increases. The right panel shows the results for $q > q^\star$. Now the zero of $\partial I_{m,fin}/\partial q$ is a maximum (as a function of $q$). The length scale $L_q$ is defined as the value of $l$ at which $\partial  I_{m,fin}/\partial q = 0$. Clearly $L_q$ decreases with increasing $q$.}
 \label{fig:LengthScale22}
 	\end{center}
 \end{figure}

 \begin{figure}[t!]
	\begin{center}
		\includegraphics[height=8cm,clip]{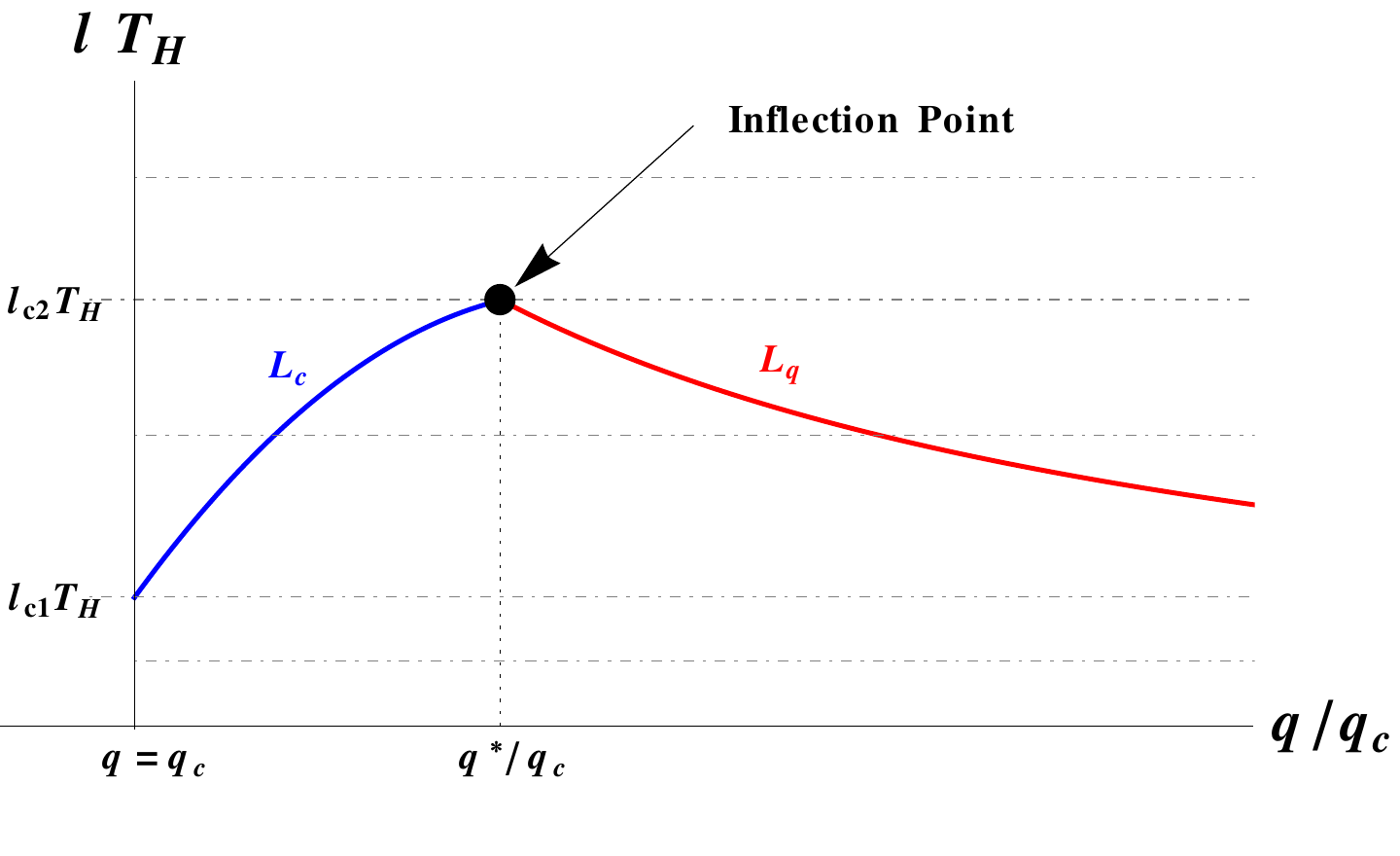}
		\caption{Phase space $(lT_H,q/q_c)$. The two curves $L_c(q/q_c)$ and $L_q(q/q_c)$ meet at $q=q^\star$. In the tent like region below these curves the derivative $(\partial I_{m,fin} / \partial q)_l >0$, outside the tent this is negative. } 
		\label{fig:LengthScale}
	\end{center}
\end{figure}

We now define two length scales $L_c, L_q$, which depend on the charge density, as follows.
\begin{eqnarray}
(\partial I_{m,fin}/\partial q )_{l=L_c} & = & 0, ~~~(\partial^2 I_{m,fin}/\partial q^2 )_{l=L_c} > 0 ~~~~~~~~q < q^\star \nonumber \\
(\partial I_{m,fin}/\partial q )_{l=L_q} & = & 0,~~~(\partial^2 I_{m,fin}/\partial q^2 )_{l=L_q}< 0 ~~~~~~~~q > q^\star
\end{eqnarray} 
From Fig (\ref{fig:LengthScale22}) it is clear that $L_c$ increases with the charge density while $L_q$ decreases with the charge density. At $q=q^\star$ one has $L_c(q^\star) = L_q(q^\star) = l_{c2}$, which is the maximum possible value of either of $L_c, L_q$.

These two length scales and their dependence on the charge density can be used to chart out a ``phase diagram'' for the entanglement entropy. Figure (\ref{fig:LengthScale}) shows the general nature of the phase diagram, parameterized by the length of a subsystem (vertical axis) and the charge density $q/q_c$  (horizontal axis).  The solid lines denote the functions $L_c(q/q_c)$ and $L_q(q/q_c)$ - as discussed above the former quantity exists for $q < q^\star$ while the latter quantity exists for $q > q^\star$. Note these curves are not real data - they are drawn to provide an impression of the general behavior. The two curves meet at $q = q^\star$, forming a "tent". At all points above this tent we have  $(\partial I_{m,fin} / \partial q)_l < 0$, while at points inside the tent we have $(\partial I_{m,fin} / \partial q)_l >0$.

It is natural to associate the scale $L_c(q)$ with the condensation and the scale $L_q(q)$ with the charge density. As remarked earlier, when the size of the subsystem is small, the entanglement entropy does not feel the effect of condensation which increases as the charge density increases. This explains why the region for small $l$ has $(\partial I_{m,fin} / \partial q)_l >0$. When $l$ is large enough, there are two competing effects - condensation tends to reduce the entanglement entropy while the charge tends to increase it. For very large $l$ the effect of condensation dominates the physics and the entanglement entropy decreases as a function of charge since increasing charge leads to more condensation. As remarked at the end of the last section, the association of $L_c(q)$ with condensation is supported by the fact that for smaller $g_{YM}$ the value of $l_{c1}$ decreases so that smaller subsystems can feel the effect of condensation. The phase diagram Fig (\ref{fig:LengthScale}) is a way to express the competition of two physical effects : the sign of $(\partial I_{m,fin} / \partial q)_l$ is a kind of ``order parameter'' which distinguishes two different physical behaviors.

It would be interesting to obtain a quantitative physical understanding of this rather novel phase diagram. 

\section{Discussions}\label{sec:discussion}

In this paper we have used holographic methods to examine how the entanglement entropy in a 1+1 dimensional field theory behaves as we change parameters such that the system crosses a critical point. The critical point in question separates a normal phase and a p-wave superconducting phase. Specifically, we studied the dependence of the entanglement entropy on the charge density for a given temperature. We found a rich behavior depending on subsystem size, with regimes of non-monotonicity. While we do not have an analytic understand of this behavior we have offered a qualitative explanation based on the competition between two opposing factors : the tendency of the entanglement to increase with increasing charge density and the depletion of degrees of freedom with increasing charge density in a condensed phase.

It would be interesting to see how the entanglement entropy behaves when we dynamically go across the critical point as in a quantum quench. In particular we would like to understand possible universal scaling of the entropy as a function of the quench rate. While some results about such scaling behavior are known in solvable and integrable field theories \cite{quenchee, cdnt} , very little is known in strongly coupled systems. In the past, holographic methods have been useful to understand scaling of one point functions in various regimes of the quench rate \cite{dreview}. The setup used in this paper should be useful in extending the discussion to entanglement entropy. 

Finally it will be interesting to explore in detail the stringy embedding of the $3d$ toy model~\eqref{GRE1}.  In the probe limit of flavor branes, the toy model~\eqref{GRE1} corresponds to a $D3-D3$ system of type IIB string theory~\cite{Gao:2012yw} without backreaction. In this limit the dilaton is a constant, which is why it is omitted in \eqref{GRE1}. In the presence of backreaction, the dilaton runs, making the problem much more difficult. It would be interesting to see if the strategy of 
\cite{Chang:2013mca} which expresses the leading order correction to the entanglement entropy in terms of the energy momentum tensors of the probe brane and the minimal surface and the bulk graviton propagator can be used to calculate this. Such a calculation would also need to include the backreaction of the four form gauge field sourced by the D3 branes. We hope to pursue this in the near future.

\acknowledgments

We would like to thank for Matteo Baggioli, Akikazu Hashimito, Song He, Matthias Kaminski, and Elias Kiritsis for helpful discussions. This work is
partially supported by the grants NSF-PHY-1521045. SRD would like to thank Tata Institute of Fundamental Research for hospitality during the completion of this paper.


\end{document}